\documentclass[journal]{IEEEtran}
\normalsize

\usepackage{amsthm}
\usepackage{graphicx}
\usepackage{verbatim}
\usepackage{amssymb,amsmath}
\usepackage{mathtools}
\usepackage[font=small]{caption}
\usepackage{epstopdf}
\usepackage{bm}
\usepackage[nosort]{cite}
\usepackage{array,color}
\usepackage[ruled,linesnumbered]{algorithm2e}
\DontPrintSemicolon
\usepackage{algpseudocode}

\newtheorem{assumption}{\textbf{Assumption}}
\newtheorem{lemma}{\textbf{Lemma}}

\newtheorem{Theorem}{\textbf{Theorem}}
\newtheorem{Remark}{\textbf{Remark}}
\usepackage[normalem]{ulem}
\usepackage{balance}
\usepackage[square,numbers]{natbib}
\usepackage{setspace}
\usepackage{subfigure}
\usepackage{multirow}
\usepackage{multicol}
\usepackage{cuted}
\usepackage{tikz} 
\usetikzlibrary{decorations.markings,arrows.meta,positioning,decorations.pathreplacing}
\tikzset{->-/.style={decoration={markings,mark=at position #1 with {\arrow{>}}},postaction={decorate}},->-/.default=0.5,
  -<-/.style={decoration={markings,mark=at position #1 with {\arrow{<}}},postaction={decorate}},-<-/.default=0.5}

\begin{document}

\title{Rateless Coding for Efficient Model Broadcast in Federated Learning}

\author{
    Zijun Qin, 
    Zuolin Jin,
    Jingxuan Huang,
    Ming Xiao,
    and Xianhao Chen
    
    \thanks{Z. Qin and X. Chen are with the Department of Electrical and Computer Engineering, University of Hong Kong, Pok Fu Lam, Hong Kong S.A.R., China (e-mail: {u3012819@connect.hku.hk}, xchen@eee.hku.hk).}
	\thanks{Z. Jin and J. Huang are with the School of Information and Electronics, Beijing Institute of Technology, Beijing, China (e-mail: zuolinjin@gmail.com, jxhbit@gmail.com).}
    \thanks{M. Xiao is with School of Electrical Engineering and Computer Science, KTH Royal Institute of Technology, Stockholm, Sweden (e-mail:mingx@kth.se).}

    \thanks{(\emph{Corresponding author: Xianhao Chen})}
}

\maketitle

{
\begin{abstract}
Model broadcast in federated learning (FL) with wireless channels faces significant challenges due to large model sizes and heterogeneous, unreliable channel conditions. While rateless coding has been widely adopted in modern broadcast communications to handle channel heterogeneity, its potential for model broadcast in FL remains largely unexplored. To fill this gap, we propose a joint source-channel coding scheme based on rateless coding to improve the robustness and efficiency of downlink model broadcast in FL. Specifically, we establish a rigorous convergence analysis of FL to inform the rateless code and quantization design under noisy downlink channel conditions. Moreover, by observing that the iterative process of FL naturally provides prior information about client updates, we leverage client-side local models as conditional prior information to aid the decoding process, thereby reducing transmission costs.
Extensive simulations demonstrate that our rateless coding scheme significantly enhances downlink model broadcast efficiency in FL, while retaining comparable training accuracies for both homogeneous and heterogeneous data distributions.
\end{abstract}


\section{Introduction}

Federated learning (FL) is a distributed machine learning paradigm that allows multiple clients to collaboratively train a global model while keeping their training data local \cite{mcmahan2017communication}. In recent years, FL has seen extensive applications in various domains, including industry \cite{9950718}, healthcare \cite{10026448}, and autonomous vehicles \cite{9827020,9945653}.
However, due to its iterative nature, FL requires frequent communications between the server and clients, which poses a significant bottleneck in practical deployments, especially in wireless networks with bandwidth-limited and unreliable channels. 

While extensive research has focused on uplink model aggregation~\cite{9210812}, efficient downlink model broadcast has received less attention. During model broadcast over a wireless downlink, the channel is often unreliable and suffers from severe degradations due to client mobility and deep channel fading, corrupting the received codewords at the clients.
Although several works have investigated efficient model broadcast in FL \cite{9448151,10669200,9515709,10210504,hu2026mixed}, they fall short in two aspects. \textit{First}, none of these schemes consider rateless coding in model broadcast. In traditional fixed-rate broadcast with feedback, such as hybrid automatic repeat request (HARQ), the transmitter must re-broadcast packets until all receivers succeed, resulting in high retransmission overhead under unreliable channels. In contrast, with rateless coding, each receiver can independently reconstruct the broadcast model as long as it accumulates enough coded symbols, substantially lowering downlink overhead across heterogeneous clients. However, despite being a predominant broadcast framework and adopted in 3GPP standards~\cite{3gpp_ts26346}, rateless coding remains unexplored in downlink FL. \textit{Second}, while the FL procedure is iterative, most of the broadcast schemes treat each downlink transmission in isolation. As a result, they fail to fully exploit (noisy) client-side local models as prior information on the server and clients, which could be leveraged to substantially reduce source entropy and downlink data volume of the global model. These observations raise two important research questions: 

\textit{\textbf{RQ1:} How can we design a rateless coding framework to achieve bandwidth-efficient downlink model broadcast in FL?}

\textit{\textbf{RQ2:} How can we further enhance it by exploiting the prior information of client local models?} 

To answer these questions, in this paper, we propose a rateless joint source-channel coding (JSCC) scheme for downlink FL based on Raptor codes, a class of rateless codes widely deployed in digital video broadcasting standards to eliminate retransmissions \cite{1-b32, 1-b33}. Specifically, we first establish a convergence bound for FL under noisy downlink broadcast to characterize the impact of quantization and decoding errors, from which we derive a tractable per-round distortion and bit error rate (BER) budget that guarantees asymptotic convergence. Guided by this theoretical foundation, we develop two core design modules: \textit{i)} to maximize communication efficiency without compromising training convergence, we jointly optimize the per-round quantization levels and Raptor encoding degree distributions for the global model weights under the derived distortion budget and client-side prior beliefs; and \textit{ii)} to construct high-confidence conditional prior probabilities of the global model at each client, we combine server-side estimations of model weight differences with client-side locally updated models, thereby accelerating belief propagation (BP) decoding and further reducing downlink overhead. The contributions of this paper are summarized as follows:
\begin{itemize}
	\item We analyze the convergence behavior of FL under noisy downlink model broadcast, based on which we establish a tractable per-round distortion budget.
	
	\item We design a rateless JSCC scheme by optimizing the quantization and coding design to minimize communication costs while preserving learning convergence.
	
	\item We exploit conditional priors from client local models to further reduce downlink bandwidth consumption.
	
	\item Simulation results show that the proposed scheme substantially reduces downlink communication bandwidth while incurring minimal performance degradation.
\end{itemize}

The remainder of this paper is organized as follows. Section II reviews the related work on communication-efficient FL. Section III introduces the system and channel model for FL with noisy downlink, as well as an overview of systematic Raptor codes. Section IV presents the convergence analysis for FL when clients initialize their local models using distorted versions of the server model. Section V investigates the joint design of coding parameters and quantization levels across training rounds, in order to optimize communication efficiency while bounding the effect of broadcast noise. Section VI presents numerical results of our proposed method. Section VII concludes our paper. The key {notations} used in our paper are summarized in Table \ref{table0}. }

\begin{table}[tp]
	\centering
	\caption{{List of important notations in this paper}}
	\label{table0}
	\begin{small}
		\setlength{\extrarowheight}{3pt}
		\begin{tabular}{|c|p{0.6\columnwidth}|}
			\hline
			Variable & Definition \\ \hline
			$K$ & Number of clients \\ \hline
			$R$ & Number of communication rounds \\ \hline
			$E$ & Number of local training steps \\ \hline
			$T$ & Total Number of local training steps \\ \hline
			$C$ & Channel capacity \\ \hline
			$Q$ & Highest quantization level \\ \hline
			$d$ & Message length \\ \hline
			$N$ & Code length \\ \hline
			$\tau$ & Transmission time slot length \\ \hline
			$\mathcal{D}$ & Model parameter size \\ \hline
			$\mathbf{u}^q_r$ & $q$-th most important bitplane at round $r$ \\ \hline
			${\mathbf{g}}^k_{r}$ & Model updates at client $k$ round $r$ \\ \hline
			$\tilde{\mathbf{g}}_{r}$ & Aggregated model updates at round $r$ \\ \hline
			$\omega^q_r$ & Degree distribution for $q$-th bitplane at round $r$ \\ \hline
			${\mathbf{w}}^k_{r, e}$ & Model weights at client $k$ round $r$ step $e$ \\ \hline
			$\tilde{\mathbf{w}}_{r}$ & Aggregated model weights at round $r$ \\ \hline
            $\hat{\mathbf{w}}^k_{r+1}$ & Server estimation for client-side model \\ \hline
			$q_r$ & Active quantization level at round $r$ \\ \hline
			$\bar D_r$ & Per-round distortion budget at round $r$ \\ \hline
			$\bar y_r$ & LDPC handoff threshold at round $r$ \\ \hline
			$\Delta_r$ & Quantization step size at round $r$ \\ \hline
			$d_{\max}$ & Maximum LT output-node degree \\ \hline
			$\epsilon_1$ & Average distortion budget over $R$ rounds \\ \hline
			$\epsilon_2$ & BER target of the error-free bitplanes \\ \hline
			$\alpha_r$ & Average variable node degree at round $r$ \\ \hline
			$P_r$ & Bit error rate at round $r$ \\ \hline
		\end{tabular}
	\end{small}
\end{table}

\section{Related Work}
\textbf{Federated learning in wireless networks.}
Many methods have been put forward to reduce the communication overhead in FL.
Some studies focus on device scheduling and resource allocation to improve the efficiency of FL.
In \cite{9210812}, a joint learning mechanism is introduced based on wireless resource allocation and user selection problem, considering the packet errors and the availability of wireless resources.
In \cite{10988901}, the authors propose a node selection scheme based on optimal stopping theory to reduce communication bandwidth and improve training accuracy.
In \cite{11314800}, an infrastructure-free federated learning framework is proposed to enable parameter mixing across large-scale geographical regions, achieving the optimal tradeoff between latency and accuracy.

Other lines of work utilize signal processing, quantization and coding techniques to reduce communication bandwidth.
In \cite{reisizadeh2020fedpaq}, the authors propose to quantize client updates before transmission and perform periodic aggregation at the server, in order to reduce communication costs.
In \cite{9878377}, the authors present a communication-efficient federated learning framework inspired by quantized compressed sensing, which greatly reduces overhead while maintaining training performance.
In \cite{9562516}, a novel over-the-air gradient aggregation scheme is proposed to approach accuracy of error-free benchmarks with limited communication bandwidth.
In \cite{9983840}, the authors perform distributed source coding of gradients using successive Wyner-Ziv coding to approach the minimum communication cost for distributed learning.

\textbf{Downlink broadcast in federated learning.}
While most research on communication-efficient FL focuses on uplink model aggregation, the downlink broadcast process has received less attention.
Several studies have investigated communication-efficient designs for downlink model transmission.
In \cite{9448151}, the authors propose an update clustering design during both uplink and downlink transmission to reduce communication overhead.
However, these schemes assume error-free downlink broadcast.
In \cite{10669200}, the authors analyze the effect of downlink decoding BER on the convergence behavior of FL, and propose an energy-efficient downlink channel coding scheme with dynamic iteration scheduling, but have not addressed the problem of reducing the length of source content.
In \cite{9515709}, the authors investigate the performance of digital and uncoded analog model broadcast over an OFDM downlink, and present a convergence analysis when heterogeneous noise patterns are present at the clients. However, the uncoded scheme could not address the problem of unreliable downlink channels where random erasures may present, and must rely on retransmission mechanisms to ensure eventual convergence, which is inefficient.
In \cite{10210504}, the authors propose a joint design of uplink and downlink communication schemes by optimizing compression, client scheduling and resource allocation. Nonetheless, the quantization scheme did not fully account for local weights at each client, which could potentially serve as side information.
In \cite{hu2026mixed}, a mixed-timescale differential coding approach is proposed for downlink model broadcast, which employs two-level differential coding for model weights to reduce bandwidth and improve resilience to decoding failures. However, the approach only considers a decoding failure probability over the entire model, which could lead to redundancy if certain parts of the model is received correctly. In \cite{9054634}, the authors mention employing rateless codes in downlink model broadcast, but did not present tractable designs that account for both robustness and efficiency.

Overall, there is a significant research gap: despite being an established and widely deployed paradigm in broadcast communications, rateless coding has not yet been exploited for model broadcast in downlink FL.

\section{System Model and Preliminaries}

In this section, we first introduce the system model of federated learning with noisy downlink broadcast. Then we provide a brief overview of Raptor codes and their decoding process.

\subsection{FL with Unreliable Downlink Broadcast}

We consider a federated learning (FL) system comprising a central server and $K$ edge clients. Each client $k \in \{1,\dots,K\}$ holds a local dataset and collaboratively trains a global model with parameters $\mathbf{w} \in \mathbb{R}^d$ under the coordination of the server. The FL objective is to minimize the global loss function:
\begin{equation}
f(\mathbf{w}) = \frac{1}{K} \sum_{k=1}^K f_k(\mathbf{w}),
\end{equation}
where $f_k(\mathbf{w})$ is the local loss function at client $k$.

The $r$-th training round consists of the following steps.

At the beginning of round $r$, each client $k$ has a local model with weights $\mathbf{w}^k_{r, 0}$. 
First, the client performs $E$ steps of stochastic gradient descent (SGD) on its local dataset:
\begin{align}
\mathbf{w}^k_{r, e+1} 
&= \mathbf{w}^k_{r, e} - \eta \mathbf{g}^k_{r, e} \notag\\
&= \mathbf{w}^k_{r, e} - \eta \nabla f_k\big(\mathbf{w}^k_{r, e}, \bm{\psi}^k_{r, e} \big), \quad e = 0,\dots,E-1,
\end{align}
where $\eta$ is the learning rate and $\nabla f_k\big(\mathbf{w}^k_{r, e}, \bm{\psi}^k_{r, e} \big)$ denotes the local gradient computed from mini-batch dataset $\bm{\psi}^k_{r, e}$. After $E$ local steps, the model update is computed as
\begin{equation}  \label{eq3}
\mathbf{g}^k_r = \mathbf{w}^k_{r,E} - \mathbf{w}^k_{r,0}.
\end{equation}

Then, each client quantizes, encodes and transmits its update $\mathbf{g}^k_r$ to the server over an uplink channel, which we assume to be error-free.
Upon receiving the client updates, the server performs averaging to obtain the aggregated update:
\begin{equation} \label{eq4}
\tilde{\mathbf{g}}_r = \frac{1}{K} \sum_{k=1}^K \mathbf{g}^k_r.
\end{equation}
The aggregated global model is then updated as
\begin{equation}  \label{eq5}
\tilde{\mathbf{w}}_{r+1} = \tilde{\mathbf{w}}_{r} + \tilde{\mathbf{g}}_r.
\end{equation}

Subsequently, the server broadcasts the aggregated model back to all $K$ clients over a downlink broadcast channel. Due to unreliable participation of edge clients and adverse wireless channel conditions, we model the downlink channel as a two-state Gilbert-Elliot channel \cite{6769369}, which has been widely considered in high-mobility scenarios \cite{5-b14, 11442636}.
At discrete time slot $j$ and client $k$, the received signal $\hat s^k_j \in \mathbb{C}^\tau$ is given by
\begin{equation}
	\label{channel_model}
    \hat s^k_j = s^k_j + n^k_j,
\end{equation}
where $s^k_j \in \mathbb{C}^\tau$ denotes the transmitted signal, and $n^k_j \sim \mathcal{CN}(0, \sigma^2_{k,ch}\mathbf{I}_\tau)$ represents the Gaussian noise. The noise power $\sigma^2_{k,ch}$ is modeled as a discrete-time Markov chain with good and bad states, i.e.
\begin{align}
    P(\sigma^2_{k,ch} &= \sigma^2_B \mid \sigma^2_{k,ch} = \sigma^2_G) = p^k, \notag\\
    P(\sigma^2_{k,ch} &= \sigma^2_G \mid \sigma^2_{k,ch} = \sigma^2_B) = q^k, \notag
\end{align}
In this work, we consider the case of severe transmission outage and assume that $\sigma^2_G \ll \sigma^2_B \to \infty$, where the bad channel state effectively leads to symbol erasures.
Note that while the server is aware of $\sigma^2_G$, it has no knowledge of $p^k$, $q^k$ at individual clients.

Finally, each client performs decoding and recovers a distorted next-round model $\mathbf{w}^k_{r+1,0}$, where the model distortion at client $k$ is denoted as
\begin{equation} \label{eq6}
{\bm{\delta}}^k_r = \mathbf{w}^k_{r+1,0} - \tilde{\mathbf{w}}_{r+1}.
\end{equation}
The training proceeds for $R$ communication rounds.

	
	

\subsection{Systematic Raptor Codes}

Traditional fixed-rate source-channel coding (i.e. LDPC codes) with retransmission mechanisms (i.e. HARQ) are inefficient under the channel model defined in Eq. (\ref{channel_model}). This is because the server has to re-broadcast the union of erased time slots to all downlink clients, where the $k$-client would also receive the erased time slots at the other $K-1$ clients, which could induce high retransmission cost given the large size of model parameters.

To solve this problem, we utilize a class of rate-adaptive codes that are resilient to random erasures named \textit{fountain codes}, or \textit{rateless codes}.
Fountain codes are a class of error-control codes designed to approximate the ``digital fountain'' appeal proposed by Byers et al.\ \cite{5-b1}, which aims to eliminate the need for retransmissions while ensuring reliable data broadcast over an erasure channel. In fountain codes, a server could generate a potentially infinite stream of coded symbols, and the receiver could reliably recover the original message upon successfully receiving \textit{any} $N$ coded symbols. The stream of coded symbols produced by the server in this solution is called a ``digital fountain''. After that, many fountain coding schemes {have} been investigated, with the most common being Luby transform (LT) codes \cite{5-b2}, online fountain codes \cite{5-b12}, spinal codes \cite{6-b16}, and their variants. 
The rateless property guarantees that, even if certain channel symbols are erased during transmission, the receiver can still recover the original message by receiving additional coded symbols from the server, even if the symbols are not identical. Hence each client does not have to receive retransmissions meant for other clients.

Raptor codes \cite{5-b3} are a class of fountain codes that achieve linear encoding and decoding complexity while approaching the channel capacity, by concatenating an outer LDPC code and an inner LT code. Specifically, we use the systematic version of raptor codes for a cleaner joint source-channel coding design. While initially meant for erasure channels, they have also been adapted for use in binary AWGN channels \cite{etesami_raptor_2006}.

To introduce the encoding and decoding process of systematic raptor codes, we make the following definitions.
Denote the original message symbols with $\mathbf{u} \in \{0,1\}^d$, the intermediate symbols with $\mathbf{\bar u} \in \{0,1\}^d$, the LDPC code symbols with $\mathbf{v} \in \{0,1\}^n$, and the LT coded symbols with $\mathbf{z} \in \{0,1\}^N$.
Define the LDPC generator matrix as $\mathbf{G}_{ldpc} \in \{0,1\}^{d \times n}$ and the corresponding parity check matrix as $\mathbf{H} \in \{0,1\}^{(n-d) \times n}$. 
Define the LT generator matrix as $\mathbf{G}_{lt} \in \{0,1\}^{n \times N}$, and a sub-matrix $\mathbf{A}_{lt} \in \{0,1\}^{n \times d}$. Each column of LT generator matrices contains $x \sim \Omega(x)$ nonzero entries whose indices are uniformly distributed in $[1, n]$.

For LT input and output nodes, denote the edge-perspective degree distributions respectively with
\begin{equation}
\iota(x)=\sum_i \iota_i x^{i-1},
\qquad
\omega(x)=\sum_j \omega_j x^{j-1}.
\end{equation}
The corresponding node-perspective distributions are
\begin{equation}
\Gamma(x)
=
\frac{\int_0^x \iota(u)\,du}
     {\int_0^1 \iota(u)\,du},
\qquad
\Omega(x)
=
\frac{\int_0^x \omega(u)\,du}
     {\int_0^1 \omega(u)\,du}.
\end{equation}
For the LDPC variable and check nodes, denote the edge-perspective degree distributions respectively with
\begin{equation}
\lambda(x)=\sum_i \lambda_i x^{i-1},
\qquad
\rho(x)=\sum_j \rho_j x^{j-1}.
\end{equation}
The node-perspective distribution of variable nodes is hence given by
\begin{equation}
\Lambda(x)
=
\frac{\int_0^x \lambda(u)\,du}
     {\int_0^1 \lambda(u)\,du}.
\end{equation}
The encoding process of systematic raptor codes is then summarized as follows.

\noindent{\textbf{Encoder:}}
\begin{enumerate}
    \item Construct $\mathbf{A}_{lt}$ so that $\mathbf{S}=\mathbf{G}_{ldpc}\mathbf{A}_{lt}$ is an invertible matrix, and $\mathbf{S}^{-1}$ is close to a random binary matrix.
    \item Encode the original message $\mathbf{u}$ into intermediate symbols with $\bar{\mathbf{u}}=\mathbf{u}\mathbf{S}^{-1}$. 
    \item Encode the intermediate symbols $\bar{\mathbf{u}}$ into LDPC coded symbols with $\mathbf{v}=\bar{\mathbf{u}}\mathbf{G}_{ldpc}$. 
    \item Encode LDPC coded symbols $\mathbf{v}$ into LT coded symbols with $\mathbf{z}=\mathbf{v}\mathbf{G}_{lt}$. 
\end{enumerate}

The decoding algorithm of Raptor codes is performed via belief propagation (BP), which can be represented with a joint factor graph. 
In the factor graph, we define three types of nodes: 
\begin{itemize}
    \item variable nodes (corresponding to $\mathbf{v}$);
    \item check nodes (corresponding to the LDPC parity checks);
    \item output nodes (corresponding to the received encoded symbols $\mathbf{z}$).
\end{itemize}
Let $m_{o,v}^{(l)}$ denote the message from an output node to a variable node at iteration $l$, and $m_{v,o}^{(l)}$ the message in the reverse direction. Similarly, $m_{c,v}^{(l)}$ and $m_{v,c}^{(l)}$ denote messages between check nodes and variable nodes. We also define virtual marginal messages between the LT and LDPC components as $m_{lt}$ and $m_{ldpc}$. Define $LLR_o$ as the channel likelihood for the LT coded symbol, given by 
\begin{equation}
	LLR_o=
	\begin{cases}
		\ln \frac{p(u_o=0)}{1 - p(u_o=0)}, \quad o = 1, 2, \ldots, d \\
		\ln \frac{p(z_{o-d}=0 \mid \hat s_{o-d})}{1 - p(z_{o-d}=0 \mid \hat s_{o-d})}, \quad o = d+1, d+2, \ldots, d+N
	\end{cases}
\end{equation}
Note that the prior belief on $u$ serves as a virtual channel to the systematic bits during decoding.
The decoding process of raptor codes can be summarized as follows.

\noindent{\textbf{Decoder:}}
\begin{enumerate}
	\item Update the messages in LT component with:
		\begin{align} \label{systematic_bp}
			\tanh \left( {{{m_{o,v}^{\left( l \right)}}}/{2}} \right) &= \tanh \left( {LLR_o/{2}} \right) \cdot \prod\limits_{v' \ne v} {\tanh \left( {{{m_{v',o}^{\left( l \right)}}}/{2}} \right)}\\
			m_{v,o}^{\left( {l + 1} \right)} &= m_{ldpc}^{\left( l \right)} + \sum\limits_{o' \ne o} {m_{o',v}^{\left( l \right)}},\\
			m_{lt}^{\left( l \right)} &= \sum\limits_{o} {m_{o,v}^{\left( l \right)}}.
		\end{align}
	\item Update the messages in LDPC component with:
		\begin{align} \label{parity_bp}
			\tanh \left( {{{m_{c,v}^{\left( l \right)}}}/{2}} \right) &= \prod\limits_{v' \ne v} {\tanh \left( {{{m_{v',c}^{\left( l \right)}}}/{2}} \right)}\\
			m_{v,c}^{\left( {l + 1} \right)} &= m_{lt}^{\left( l \right)} + \sum\limits_{c' \ne c} {m_{c',v}^{\left( l \right)}},\\
			m_{ldpc}^{\left( l \right)} &= \sum\limits_{c} {m_{c,v}^{\left( l \right)}}.
		\end{align}
	\item Update the messages iteratively until convergence.
\end{enumerate}

\section{Convergence Analysis of FL with Noisy Downlink Broadcast}

In this section, we analyze the convergence behavior of federated learning when
clients initialize their local models using distorted versions of the server
model. 

We impose the following assumptions, following existing literature on distributed optimization \cite{zhang2012communication} \cite{karimireddy2020scaffold}.
\begin{assumption}[Smoothness]
\label{ass:smoothness}
The global objective $f(\cdot)$ and local objectives $f_k(\cdot)$ are $L$-smooth:
\begin{align}
	\|\nabla f(\mathbf{w})-\nabla f(\mathbf{w'})\|
    &\le L\|\mathbf{w}-\mathbf{w'}\|, \notag\\
    \|\nabla f_k(\mathbf{w})-\nabla f_k(\mathbf{w'})\|
    &\le L\|\mathbf{w}-\mathbf{w'}\|,
    \qquad \forall\,\mathbf{w},\mathbf{w'}.
\end{align}
\end{assumption}


\begin{assumption}[Unbiased stochastic gradients]
\label{ass:unbiased_gradients}
Let $\mathcal{F}_{r,e}$ denote the history before local step $e$ of round
$r$. The stochastic gradient $\mathbf{g}_{r,e}^k$ satisfies
\begin{equation}
    \mathbb{E}\!
    \left[\mathbf{g}_{r,e}^k\mid\mathcal{F}_{r,e}\right]
    =\nabla f_k(\mathbf{w}_{r,e}^k).
\end{equation}
\end{assumption}

\begin{assumption}[Bounded stochastic-gradient variance]
\label{ass:bounded_local_variance}
For every $r,e$, and $k$,
\begin{equation}
    \mathbb{E}\!\left[
    \|\mathbf{g}_{r,e}^k-\nabla f_k(\mathbf{w}_{r,e}^k)\|^2
    \mid \mathcal{F}_{r,e}\right]
    \le \sigma_L^2.
    \label{eq:local_gradient_variance_assumption}
\end{equation}
The stochastic-gradient noises are conditionally uncorrelated across clients:
\begin{align}
&\mathbb{E}\!\left[
\left\langle
\mathbf{g}_{r,e}^k-\nabla f_k(\mathbf{w}_{r,e}^k),
\mathbf{g}_{r,e}^{\ell}-\nabla f_{\ell}(\mathbf{w}_{r,e}^{\ell})
\right\rangle
\middle|\mathcal{F}_{r,e}\right]=0,
\quad k\ne\ell.
\label{eq:cross_client_noise_assumption}
\end{align}
\end{assumption}

\begin{assumption}[Bounded gradient dissimilarity]
\label{ass:gradient_dissimilarity}
There exists $\sigma_G^2\ge0$ such that
\begin{equation}
    \frac{1}{K}\sum_{k=1}^{K}
    \|\nabla f_k(\mathbf{w})-\nabla f(\mathbf{w})\|^2
    \le \sigma_G^2,
    \qquad \forall\,\mathbf{w}.
    \label{eq:gradient_dissimilarity_assumption}
\end{equation}
\end{assumption}

Define a virtual server model for each local step $e$ of round $r$ as
\begin{equation}
    \widetilde{\mathbf{w}}_{r,e}
    \triangleq
    \widetilde{\mathbf{w}}_r
    -\eta\sum_{s=0}^{e-1}\overline{\mathbf{g}}_{r,s},
    \label{eq:virtual_server_iterate}
\end{equation}
where
\begin{equation}
    \overline{\mathbf{g}}_{r,e}
    \triangleq \frac{1}{K}\sum_{k=1}^{K}\mathbf{g}_{r,e}^k,
\end{equation}
and
\begin{equation}
    \widetilde{\mathbf{w}}_{r,e+1}
    =\widetilde{\mathbf{w}}_{r,e}
    -\eta\overline{\mathbf{g}}_{r,e}.
    \label{eq:virtual_server_update}
\end{equation}

The bound on one step of gradient descent is given in the following lemma.
\begin{lemma}\label{lemma1}
The expected value of the global objective after one local step satisfies
\begin{align}
&\mathbb{E}f(\widetilde{\mathbf{w}}_{r,e+1})
\le \mathbb{E}f(\widetilde{\mathbf{w}}_{r,e})
-\frac{\eta}{2}
 \mathbb{E}\|\nabla f(\widetilde{\mathbf{w}}_{r,e})\|^2
\notag\\
&\quad+\frac{\eta L^2}{2}\frac{1}{K}\sum_{k=1}^{K}
    \mathbb{E}\|\mathbf{w}_{r,e}^k-
    \widetilde{\mathbf{w}}_{r,e}\|^2
+\frac{L\eta^2\sigma_L^2}{2K}.
\label{eq:one_step_descent_corrected}
\end{align}
\end{lemma}

The proof is deferred to Appendix \ref{app:proof_lemma1}.

The following lemma provides an upper bound on the disagreement between the local models and the virtual server model.
\begin{lemma} \label{lemma2}
The overall drift of the local models from the virtual server model is upper-bounded by
\begin{align}
&\sum_{e=0}^{E-1}\frac{1}{K}\sum_{k=1}^{K}
    \mathbb{E}\|\mathbf{w}_{r,e}^k-
    \widetilde{\mathbf{w}}_{r,e}\|^2 \le4E\frac{1}{K}\sum_{k=1}^{K} \mathbb{E}\|\bm{\delta}_{r-1}^k\|^2 \notag\\
&+12\eta^2E^2\sum_{e=0}^{E-1}
\mathbb{E}\|\nabla f(\widetilde{\mathbf{w}}_{r,e})\|^2
+4\eta^2E^3(3\sigma_G^2+\sigma_L^2).
\label{eq:summed_drift_bound}
\end{align}
\end{lemma}

The proof is deferred to Appendix \ref{app:proof_lemma2}.

Combining Lemma \ref{lemma1} and Lemma \ref{lemma2}, we obtain the following convergence result for FL with noisy downlink broadcast.
\begin{Theorem}[Convergence with noisy downlink initialization]
\label{Theorem1}
If
\begin{equation}
    \eta \le\frac{1}{\sqrt{24}LE},
    \label{eq:theorem_stepsize_condition}
\end{equation}
then, after $R$ communication rounds and $E$ local updates per round,
\begin{align}
&\frac{1}{RE}\sum_{r=0}^{R-1}\sum_{e=0}^{E-1}
\mathbb{E}\|\nabla f(\widetilde{\mathbf{w}}_{r,e})\|^2
\le
\frac{4\left(f(\widetilde{\mathbf{w}}_0)-f^\star\right)}
{\eta RE} \notag\\
&+\frac{8L^2}{R}\sum_{r=0}^{R-1}\frac{1}{K}\sum_{k=1}^{K} \mathbb{E}\|\bm{\delta}_{r-1}^k\|^2
+8\eta^2L^2E^2(3\sigma_G^2+\sigma_L^2)
+\frac{2L\eta\sigma_L^2}{K}.
\label{eq:theorem1_corrected_bound}
\end{align}
\end{Theorem}

The proof is deferred to Appendix \ref{app:proof_theorem1}.

From Theorem \ref{Theorem1}, we can see that the effect of downlink distortion could be captured by a distortion term $D_r$ defined as
\begin{equation} \label{eq54}
	D_r\triangleq\frac{8L^2}{K}\sum_{k=1}^{K} \mathbb{E}\|\bm{\delta}_{r-1}^k\|^2.
\end{equation}
This term is proportional to the expected square distortion, accumulates with training rounds $R$, and is not affected by learning rate $\eta$.
The following conditions on asymptotic convergence can be immediately derived.
\begin{Remark}[Asymptotic convergence]
\label{corollary1}
Let $E$ be fixed and choose
$\eta=\frac{c}{\sqrt{RE}}$,
where $c>0$ is selected so that
\eqref{eq:theorem_stepsize_condition} holds. If
\begin{equation}
    \lim_{R\to\infty}\frac{1}{R}
    \sum_{r=0}^{R-1}D_r=0,
    \label{eq:vanishing_average_distortion_condition}
\end{equation}
then the average squared gradient norm converges to 0, and the dominant terms decay as
\begin{equation}
    \lim_{R\to\infty}
    \frac{1}{RE}\sum_{r=0}^{R-1}\sum_{e=0}^{E-1}
    \mathbb{E}\|\nabla f(\widetilde{\mathbf{w}}_{r,e})\|^2 \le O\!\left(\frac{1}{\sqrt{RE}}\right) + \frac{1}{R}\sum_{r=0}^{R-1}D_r. \notag
\end{equation}
\end{Remark}

\section{Rateless Downlink Model Broadcast for Federated Learning}

In this section we present the schematic design of our proposed rateless downlink broadcast scheme for federated learning. First we present the overall design of rateless downlink model broadcast in FL. Then we introduce the extrinsic information transfer (EXIT) chart analysis of systematic Raptor codes in the presence of prior beliefs for the source. Finally we optimize system parameters for our proposed scheme to reduce bandwidth consumption while ensuring convergence.

\subsection{System Overview}
\begin{figure*}[ht]
	\setlength{\belowcaptionskip}{-0.5cm}
	\vspace{-20pt}
	\centering
	\includegraphics[width=17cm]{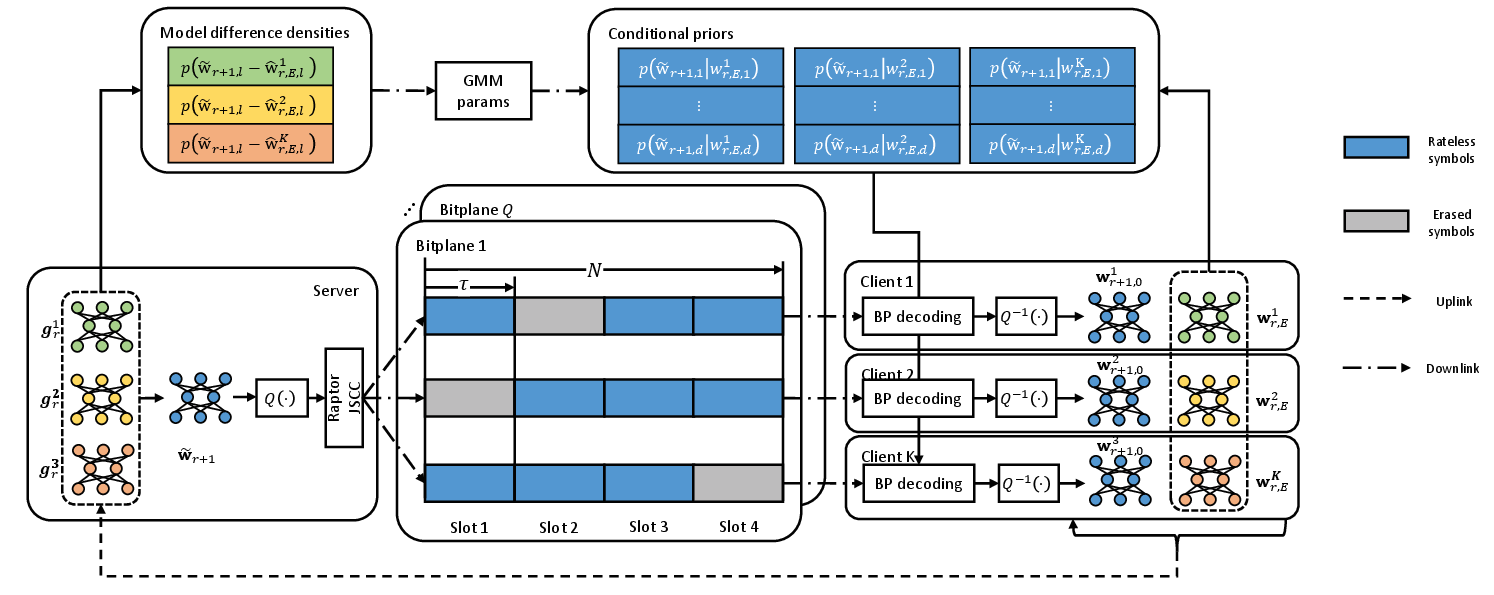}
	\caption{Schematic diagram for rateless model broadcast.}
	\label{fig2}
\end{figure*}
In this part, we describe our proposed downlink transmission procedure for efficient model broadcast. The overall scheme is depicted in Fig. \ref{fig2}.

In the $r$-th training round, the server performs aggregation of client updates to obtain the aggregated model $\tilde{\mathbf{w}}_{r+1}$. After that, the server performs the following steps.

\noindent \textbf{(1) Model difference density estimation.}

The server estimates the updated weights at each client using
\begin{equation}
	\hat{\mathbf{w}}^k_{r+1} \triangleq  \tilde{\mathbf{w}}_{r} + \mathbf{g}^k_r,
\end{equation}
then the prediction of the client weight changes is defined as
\begin{equation} \label{eq66}
	\hat{\mathbf{z}}^k_{r} \triangleq \tilde{\mathbf{w}}_{r+1} - \hat{\mathbf{w}}^k_{r+1} = \tilde{\mathbf{g}}_r - \mathbf{g}^k_r.
\end{equation}
By assuming $\hat{\mathbf{z}}^k_{r}$ as an i.i.d random variable, the server is able to estimate $p(\hat{{z}}^k_{r})$ using $\tilde{\mathbf{g}}_r$ and $\mathbf{g}^k_r$. 
In our proposed system, we model the density of predicted model difference as a Gaussian mixture model (GMM) which is i.i.d. across model weights, and approximate the difference at each position using $M$ components, given by
\begin{equation} \label{eq67}
	p(\hat{{z}}^k_{r}) = \sum_{m=1}^{M} \pi^k_m \mathcal{N}(\hat{{z}}^k_{r} \mid \mu^k_m, (\sigma^k_m)^2), \quad l = 1,2,\ldots, d.
\end{equation}
where $\pi^k_m$, $\mu^k_m$ and $(\sigma^k_m)^2$ are the mixture weight, mean, and variance of the $m$-th Gaussian component, respectively, and are determined with the expectation maximization algorithm. The GMM parameters are then sent to the clients with negligible overhead.

\noindent \textbf{(2) Quantization.}

To reduce the communication overhead, the server first quantizes $\tilde{\mathbf{w}}_{r+1}$ to $q_r$ bits per weight using a uniform quantizer. Denote the quantization range as 
\begin{equation}
    M_r \triangleq T_{\max,r} - T_{\min,r},
\end{equation}
where $T_{\min,r}$ and $T_{\max,r}$ are the minimum and maximum values of quantization thresholds at round $r$. 
The edges of the $n$-th quantization bin of the active quantizer are given by
\begin{equation}
    T_{n,r} = T_{\min,r} + n\Delta_r, \quad n = 0,\dots,2^{q_r},
\end{equation}
 the quantization step is defined as
\begin{equation}
	\Delta_r = M_r / \left(2^{q_r}-1\right),
\end{equation}
where $q_r \in \{1,\dots,Q\}$ denotes the active quantization level of round $r$. The quantizer employs \emph{subtractive dithering}: the server draws an i.i.d.\ uniform noise 
\[\xi_{r,l}\sim\mathcal{U}[-0.5, 0.5]\]
that is independent of the model weights, adds it prior to rounding, and the clients subtract the same realization after decoding, i.e.,
\begin{equation} \label{eq:quantization}
\text{Quantize}\left(\tilde{w}_{r,l}\right) = \text{round}\left[\frac{\tilde{w}_{r,l} - T_{\min,r}}{\Delta_r} + \xi_{r,l}\right],
\end{equation}
and the $i$-th quantized bitplane is subsequently calculated with
\begin{equation} \label{eq_binarize}
    u^i_{r,l}=\left\lfloor \frac{\text{Quantize}\left(\tilde{w}_{r,l}\right)}{2^{q_r-i}} \right\rfloor \bmod 2, \quad i = 1,\dots,q_r.
\end{equation}
The dither is generated from a per-round pseudo-random seed that is announced in the broadcast, so it is available at both ends and costs negligible bandwidth.

\noindent \textbf{(3) Raptor JSCC.}

Each transmitted bitplane is then encoded using a Raptor code and sent over a binary-input additive white Gaussian noise (BIAWGN) channel using BPSK modulation. 
For each quantization level, we divide each bitplane into blocks of length $d$. For each block, we apply a raptor joint source-channel code to produce streams of rateless channel symbols, where the successful reception of $N^{q_r}_r$ channel symbols could achieve the desired BER level.
Specifically, at round $r$, the active quantization level is denoted by $q_r \in \{1,\dots,Q\}$, for which we assume the following:
\begin{itemize}
	\item The $q_r-1$ most significant bitplanes $\mathbf{u}^1_r,\dots,\mathbf{u}^{q_r-1}_r$ are transmitted with very low BER, and can be seen as error-free during analysis.
	\item The least significant transmitted bitplane $\mathbf{u}^{q_r}_r$ is transmitted with non-negligible BER $P_r$.
	\item The remaining lower bitplanes are not transmitted and are treated as all-zero bitplanes at the clients.
\end{itemize}

\noindent \textbf{(4) BP decoding.}

Based on the received GMM parameters, each client produces an estimation for the density of $\tilde{{w}}_{r+1, l}$ based on ${{w}}^k_{r, E, l}$, denoted with
\[\hat p(\tilde{{w}}_{r+1, l} \mid {{w}}^k_{r, E, l}).\]
The density is then utilized to calculate bitplane prior probabilities
\[p\!\left(u_{r+1, l}^q = b \mid \{u_{r+1, l}\}_1^{q-1}, {{w}}^k_{r, E, l}\right).\]
The bitplanes are decoded with both the prior probability for systematic bits using Eq. (\ref{systematic_bp}), and the parity symbol likelihoods conveyed via the rateless symbols using Eq. (\ref{parity_bp}).
  with $\left\{\hat{\mathbf{u}}^i_r\right\}_{i=1}^{q_r}$.

\noindent \textbf{(5) Dequantization.}

Let 
\[\nu_{r+1, l}^{q_r} \triangleq \sum_{i=1}^{q_r} \hat u^i_{r+1,l} 2^{q_r-i}\]
denote the integer represented by the $q_r$ most significant bits $\{u^i_{r,l}\}_{i=1}^{q_r}$, and the dequantization operation is defined as 
\begin{equation} \label{eq:dedither}
w^k_{r+1,0,l}=\text{Dequantize}\left(\nu_{r+1, l}^{q_r}\right) = T_{\min,r} + (\nu_{r+1, l}^{q_r} - \xi_{r,l})\,\Delta_r.
\end{equation}

The overall design objective is to minimize the bandwidth consumption of worst-case client while guaranteeing the eventual convergence of the FL process, then the problem is formulated as:
\begin{align}
\textbf{P1:} \quad \min_{\{q_r, \alpha_r, \omega^i_r, \bar y_r\}}
& \sum_{i=1}^{q_r}N^i_r \\
\text{s.t.} \quad
& \sum_{j=1}^{d_{max}} \omega^i_{r, j} = 1, \label{eq32}\\
& BER_i \le \epsilon_2, \quad i = 1,\dots,q_r-1 \label{eq33.1}\\
& \frac{1}{RK} \sum_{r=1}^{R} \sum_{k=1}^{K} D^k_r \le \epsilon_1, \label{eq33}\\
& \lim_{R \to \infty} \frac{1}{RE} \sum_{r=1}^{R} \sum_{e=0}^{E-1} \mathbb{E}\left\|\nabla f(\widetilde{\mathbf{w}}_{r,e})\right\|^2 = 0, \label{eq34}\\
& q_r \in \{1,\dots,Q\}, \notag\\
& r = 1,\dots,R, \quad l = 1,\dots,\mathcal{D}. \notag
\end{align}
In this problem, constraint (\ref{eq32}) ensures that the degree distribution is valid, constraint (\ref{eq33.1}) bounds the BER of previous bitplanes below a threshold $\epsilon_2$, constraint (\ref{eq33}) bounds the impact of the noise term to a pre-determined level $\epsilon_1$ after $R$ communication rounds, and constraint (\ref{eq34}) ensures that the FL process converges to a stationary point as $R \to +\infty$, $\Delta_R \to 0$.

However, problem \textbf{P1} is difficult to solve directly for several reasons. First, $D^k_r$ is dependent on the distribution of model weights in each individual round, hence a unified design across $r$ is not tractable. Secondly, both the quantization level $q_r$ and the degree distribution $\omega^i_{r, j}$ affect the objective function. Finally, the bit error rate for iterative BP decoding can only be analyzed recursively. 
Therefore, in the remainder of this section, we investigate how to decompose \textbf{P1} into several sub-problems.

\subsection{Per-round Quantization and BER Budget}
In this part, we simplify constraints (\ref{eq33}) and (\ref{eq34}) by deriving a per-round quantization and BER budget that serves a sufficient condition for convergence of FL with noisy downlink broadcast.

Following Remark \ref{corollary1}, we introduce a per-round distortion budget to facilitate asymptotic convergence, given by
\begin{equation} \label{eq64}
D^k_r \le \frac{D_0}{r + \beta_0}, \quad \beta_0 > 0.
\end{equation}
This budget serves to satisfy constraint Eq. (\ref{eq34}), where the average distortion approaches zero asymptotically, given by
\begin{equation}
    \frac{1}{R}\sum_{r=1}^{R}D_r
    \le\frac{D_0}{R}\sum_{r=1}^{R}\frac{1}{r+\beta_0}
    =O\!\left(\frac{\log R}{R}\right),
\end{equation}
which goes to 0 when $R$ approaches infinity.
Moreover, substituting Eq. (\ref{eq64}) into Eq. (\ref{eq33}) and solving for $\beta_0$ yields
\begin{equation}
\beta_0 \ge \frac{R}{e^{\epsilon_1 R/D_0} - 1}.
\label{eq58}
\end{equation}
By taking the equality, we define a per-round distortion budget as
\begin{equation} \label{distortion_budget}
{D^k_r}\le \bar D_r \triangleq \frac{D_0}{r + \frac{R}{e^{\epsilon_1 R/D_0} - 1}}.
\end{equation}

Next, we derive the relationship between the distortion term $D^k_r$ and the quantization level $q_r$ and target error rate $P_r$ of the $q_r$-th bitplane.
To simplify design, we assume the same quantization level $q_r$ and target error rate $P_r$ for all clients. Therefore, the distortions at the clients can be modeled as i.i.d. random variables with the same distribution, as follows.
\begin{lemma} \label{distortion_distribution}
	Define $p({\delta}_r)$ as the probability density function of the client distortion at a single model coordinate. 
	With the subtractive dithering of Eq. (\ref{eq:quantization}) and a bit error on the active bitplane, for $q_r \le Q$, the overall distortion is given by
    \begin{equation}
        \delta_{r-1} = e_r + \Delta_r(1 - 2u^{q_r}_r)F,
    \end{equation}
    where $e_r \sim\mathcal{U}[-0.5, 0.5]$ is the quantization error, $F \sim \text{Bernoulli}(P_r)$ indicates decoding error of the active bitplane.
\end{lemma}

\begin{Theorem} \label{theorem2}
The distortion term could be characterized by the quantization level $q_r$, the target error rate $P_r$ and the quantization range $M_r$
\begin{equation}
    D_r(P_r,q_r,M_r)
    \le 8L^2\mathcal{D}\left(\frac{1}{12} + 2P_r\right)\Delta^2_r,
\end{equation}
where $\mathcal{D}$ is the dimension of the model and $D_r=D^k_r$ holds under the i.i.d. assumption on the per-client distortions.
\end{Theorem}
\begin{proof}
    From Lemma \ref{distortion_distribution}, the per-coordinate second moment of distortion is
    \begin{equation} \label{eq30}
    \mathbb{E}\!\left[\delta_r^2\right]
    =\left(\frac{1}{12} + P_r\right)\Delta^2_r + 2\Delta\mathbb{E}\left[e_r(1 - 2u^{q_r}_r)F\right].
    \end{equation}
    Since $e_r$ is not independent from $2u^{q_r}_r$, we bound the final term as follows
    \begin{equation}
        \mathbb{E}\left[e_r(1 - 2u^{q_r}_r)F\right] \le |\mathbb{E}\left[e_r(1 - 2u^{q_r}_r)F\right]| \le |\mathbb{E}[|e_r|F] \le \frac{\Delta_r}{2}P_r,\notag
    \end{equation}
    which leads to 
    \begin{equation}
        \mathbb{E}\!\left[\delta_r^2\right] \le \left(\frac{1}{12} + 2P_r\right)\Delta^2_r.
    \end{equation}
Since $\mathbb{E}\|\bm{\delta}_r\|^2 = \mathcal{D}\,\mathbb{E}[\delta_r^2]$ for a $\mathcal{D}$-dimensional model, the proof is complete.
\end{proof}

The distortion term in Eq. \eqref{eq54} can be directly linked to the constraints of quantization and coding design.

Combining Theorem \ref{theorem2} and Eq. (\ref{distortion_budget}), we can obtain the target error rate for each round.
\begin{Remark}
\label{corollary:p_max}
	For given quantization level $q_r$ and current quantization range $M_r$, the target error rate $P_r$ is given by
	\begin{equation} \label{ber_target}
		P_r \triangleq \frac{1}{2}\left(\frac{\bar D_r}{8L^2\mathcal{D}\Delta^2_r} - \frac{1}{12}\right).
	\end{equation}
\end{Remark}

This allows us to ensure the constraints Eq. (\ref{eq33}) (\ref{eq34}) with a simpler per-round distortion budget that guarantees bounded noise behavior of finite rounds, as well as asymptotic convergence of FL.

\subsection{Multilayer EXIT Chart Analysis for Raptor-Based JSCC}
In this part, we present the multilayer extrinsic information transfer (EXIT) analysis \cite{957394} of systematic raptor codes, in order to convert BER constraints in Eq. (\ref{eq33.1}) to linear constraints.

Under the conditions of binary symmetric channel, BPSK modulation, and high node degrees, every BP message could be approximated with a symmetric Gaussian random variable
\begin{equation}
m \sim \mathcal{N}(\mu,2\mu),
\end{equation}
which is completely characterized by its mean $\mu$. 
Therefore, the mutual information between a binary variable $U\in\{0,1\}$ and the message $m$ is
\begin{equation}
I(U;m) = 1-\mathbb{E} \left[\log_2\!\left(1+e^{-m}\right)\right] \triangleq J(\mu).
\end{equation}
It can be verified that $J(\mu)$ is a monotonically increasing function of $\mu$, with $J(0)=0$ and $\lim_{\mu\to\infty} J(\mu)=1$.

We introduce the following definitions:
\begin{align}
	x = J(\mu_{v,o}), \quad
	y = J(\mu_{o,v}), \quad
	X = J(\mu_{v,c}), \quad
	Y = J(\mu_{c,v}). \notag
\end{align}
Moreover, define $\beta \triangleq {d}/(d+N)$, where $d$ is the number of original message symbols and $N$ is the number of LT coded symbols. Therefore, a fraction $\beta$ of LT output nodes corresponds to source-symbol observations and the remaining fraction $1-\beta$ corresponds to physical channel observations of capacity $C$. 
In \cite{lossy_raptor}, the authors derive the LT EXIT recursion as
\begin{equation} \label{eq76}
x^{(t+1)}=\sum_k \sum_i\Lambda_k \iota_i
J\!\left((i-1)J^{-1}(y^{(t)})+kJ^{-1}(Y^{(t)})\right),
\end{equation}
and
\begin{align} \label{eq77}
y^{(t+1)}&=1- \sum_j \omega_j
\Big[\beta J \!\left((j-1)J^{-1}(1-x^{(t)})+J^{-1}(H)\right) \notag\\
&+(1-\beta)J\!\left(
(j-1)J^{-1}(1-x^{(t)})+J^{-1}(1-C)
\right)
\Big].
\end{align}
Here, $H$ denotes the binary entropy of the systematic output node.
The LDPC precode EXIT recursion is
\begin{equation} \label{eq78}
X^{(t+1)}=\sum_k \sum_i\lambda_k \Gamma_iJ\!\left((k-1)J^{-1}(Y^{(t)})+iJ^{-1}(y^{(t)})\right),
\end{equation}
and
\begin{equation} \label{eq79}
Y^{(t+1)}=1-\sum_j \rho_jJ\!\left((j-1)J^{-1}(1-X^{(t)})\right).
\end{equation}

\begin{figure*}
\vspace{-20pt}
\begin{align} \label{y_multilayer}
y=1-\sum_j \omega_j\Bigg[
(1-\beta)
J\!\left((j-1)J^{-1}(1-x)+J^{-1}(1 - C)\right) +\beta\sum_{\{u\}_1^{q-1}}p(\{u\}_1^{q-1})J\!\left((j-1)J^{-1}(1-x)+J^{-1}\!\left(H({u}^q|\{u\}_1^{q-1})\right)\right)
\Bigg]
\end{align}
\vspace{-20pt}
\begin{align} \label{eq84}
h_j(y) &= 
{\beta}\sum_{\{u\}_1^{q-1}}p(\{u\}_1^{q-1})J\!\Bigg[(j-1)J^{-1}\left(1-\sum_i \iota_iJ\!\left((i-1)J^{-1}(y)\right)\right)+J^{-1}\!\left(H({u}^q|\{u\}_1^{q-1})\right)\Bigg] \notag\\
&+(1-\beta)J\!\Bigg[
(j-1)J^{-1}(1-\sum_i
\iota_i
J\!\left((i-1)J^{-1}(y)\right))+J^{-1}(1-C)\Bigg]. 
\end{align}
\hrulefill
\end{figure*}

However, due to the nonlinear coupling between the LT and LDPC components, we must further modify the decoding procedure to arrive at a tractable linear program over the LT degree distribution.
Following the approach in \cite{lossy_raptor}, we decompose the joint LT-LDPC decoding into two separate stages, and define a handoff point $\bar y$ at which the LT decoder stops and the LDPC decoder starts.
First we perform LT decoding iterations while ignoring the LDPC feedback by setting $Y=0$ in Eq. (\ref{eq76}), until a threshold $\bar y$ is reached. The update to $y$ in each iteration could be simplified as 
\begin{equation}
y^{(t+1)}=1-\sum_j \omega_jh_j(y^{(t)}),
\end{equation}
where $h_j(\cdot)$ is given by Eq. (\ref{eq84}).
Then we perform LDPC decoding iterations while ignoring the LT feedback by setting $y=\bar y$ in Eq. (\ref{eq78}), and the update to $Y$ in each iteration could be simplified as
\begin{align} \label{IVC_ldpc}
Y^{(t+1)}&=1-\sum_j \rho_jJ\Bigg[(j-1)J^{-1}\Bigg(1-\sum_k \sum_i\lambda_k \Gamma_i \notag\\
&\quad J\!\left((k-1)J^{-1}(Y^{(t)})+iJ^{-1}(\bar y)\right)\Bigg)\Bigg].
\end{align}
The LDPC recursion \eqref{IVC_ldpc} is run to convergence at the handoff point $y = \bar y$, which yields the final LDPC extrinsic value $\bar Y$, which is returned into LT decoding.
The final operating point $\bar x$ is then obtained from the LT variable-node recursion using
\begin{equation}
\bar x
=
\sum_k \sum_i
\Lambda_k \iota_i
J
\!\left(
(i-1)J^{-1}(\bar y)
+
kJ^{-1}(\bar Y)
\right).
\end{equation}
Consider the $q$-th most significant bitplane, let $H({u}^q|\{u\}_1^{q-1})$ denote the entropy of the current bitplane for a message symbol, conditioned on the $q-1$ more-significant bitplanes that have already been recovered without error. 
There are $2^{q-1}$ possible conditioning states, with probabilities $p(\{u_l\}_1^{q-1})$. 
The LT check-node update must average over all conditional source states, and the resulting multilayer EXIT equation becomes Eq. (\ref{y_multilayer}).
The asymptotic bit error rate (BER) of the decoded bitplane can be estimated from the output LLR mean. The output LLR for a degree-$j$ source symbol is given by
\begin{align}
\bar\mu_j(\bar y)&=J^{-1}\biggl\{1 - J\left[jJ^{-1}(1- \bar x)\right]\biggr\} \notag\\
&+ \sum_{\{u\}_1^{q-1}}p(\{u\}_1^{q-1})J^{-1}\!\left(1 - H({u}^q|\{u\}_1^{q-1})\right).
\end{align}
Note that $\bar\mu_j(\bar y)$ is entirely determined by $\bar y$ when the LDPC variable node degree distribution $\Lambda(x)$, check node distribution $\Gamma(x)$, and the average LT variable node degree $\alpha$ are given. In our implementation, the LT input-node degree distribution is taken to be a Poisson distribution with mean $\alpha$.

For a target quantization level \(q\), we define two inner problems. 
The first inner problem is optimizing the LT degree distribution for the currently active quantization level $q$, in order to minimize the feasible coding overhead while constrained by its target BER $P_r$.
Define the per-degree BER basis as \[P_j(\bar y) = Q
\!\left( \sqrt{\frac{\bar \mu_j(\bar y)}{2}} \right).\]
then the BER is calculated by $P=\sum_j \Omega_j P_j(\bar y),$
where $\Omega_j$ is the node-perspective LT output-node degree distribution and $Q(\cdot)$ is the decoding error function.

\subsection{Code Optimization Design}

In this part we decompose the original problem \textbf{P1} into an outer quantization-level optimization problem and inner degree distribution optimization problems, and propose a practical algorithm to solve those problems. 

The EXIT analysis allows us to convert the design of Raptor codes into a linear program with respect to $\omega$. For a fixed LDPC code, the optimization of LT degree distribution with respect to a fixed $\alpha$, $\bar y$ and a target BER $P$ is formulated as
\begin{align} 
\textbf{P2:} \quad \Phi_r(q) & \triangleq \min_{\{\alpha_r, \omega^q_r, \bar y_r\}} \quad \alpha_r\sum_{j=1}^{d_{\max}}\frac{\omega^q_{r, j}}{j}-1 \\ 
\mathrm{s.t.} 
\quad 
& 1-\sum_{j=1}^{d_{\max}}\omega^q_{r, j}h_j(y) > y + \epsilon_3, \quad y \in (0, \bar y_r] \label{eq90}\\ 
& \sum_{j=1}^{d_{\max}}\frac{\omega^q_{r, j}}{j}\left[P_{r, j}(\bar y) - P_r\right] < 0, \label{eq91}\\
& \sum_{j=1}^{d_{\max}}\omega^q_{r, j}=1, \\ 
& \omega^q_{r, j}\ge 0.
\end{align}
Here, the objective is the per-block overhead $N^q_r$ reformulated using the identity
\[\frac{N^q_r}{d} = \frac{\alpha_r}{\sum^{d_{max}}_{j=1}j\Omega^q_{r,j}}-1 = \alpha_r\sum_{j=1}^{d_{\max}}\frac{\omega^q_{r, j}}{j}-1.\]
Constraint Eq. (\ref{eq90}) ensures that the LT EXIT curves are separated before reaching the threshold $\bar y$, indicating that the iterative decoding of LT component could progress until feeding the decoded LLR to the LDPC component. Constraint Eq. (\ref{eq91}) ensures that the final BER of the decoded bitplane is below the target threshold. Both constraints are evaluated at the operating points determined by the decomposition above: the EXIT tunnel \eqref{eq90} is enforced on a discrete grid of $y \in (0, \bar y]$, and, because the decoded BER is linear in the node-perspective distribution $\Omega_j$, the BER constraint \eqref{eq91} is implemented as a single equivalent linear row in the edge-perspective variables $\omega_j$ evaluated at the final operating point $\bar x$. Note that while the value of $\beta$ depends on $\omega^q_{r,j}$, we decouple them by heuristically setting it to ${C}/\left[C + H({u}^q|\{u\}_1^{q-1})\right]$, following the method in \cite{lossy_raptor}.

\textbf{P2} can be solved by a linear program with respect to $\omega^q_r$ for fixed $\alpha_r$ and $\bar y_r$, together with a grid search over $\alpha_r$ and $\bar y_r$ to find the optimal solution. During optimization, we sample discrete values for $y$ in the interval $(0, \bar y_r]$ and treat Eq. (\ref{eq90}) as a finite set of linear constraints. For each candidate pair $(\alpha_r, \bar y_r)$, the LT input-node degree distribution is set to a Poisson distribution with mean $\alpha_r$, the LDPC precode is fixed to a regular profile (degree-$3$ variable nodes and degree-$20$ check nodes), and the final operating point $\bar x$ together with the per-degree BER basis is evaluated as described in the previous subsection. Because the BER constraint \eqref{eq91} is linear in the node-perspective distribution, it is rewritten as a single linear row in the edge-perspective variables $\omega^q_r$, and \textbf{P2} is solved with a standard linear-programming solver. 

For the $i$-th bitplane where $i < q$, the inner problem minimizes the coding overhead while constrained by the target bit error rate $\epsilon_2$. Since these bitplanes must be decoded reliably, we reuse the same LP formulation as \textbf{P2} with the BER target replaced by $\epsilon_2$, leading to \textbf{P2'}.
\textbf{P2'} can be solved by reusing the solver for \textbf{P2} with the BER target replaced by $\epsilon_2$.

Define the outer problem as optimizing quantization-level to minimize the total minimum-achievable coding overhead across chosen quantization levels, given by
\begin{align} 
\textbf{P3:} \quad \min_{q_r\in\{1,\dots,Q\}} \quad & \Psi_r(q_r - 1) + \Phi_r(q_r). 
\end{align}
Here, define
\begin{equation}
	\Psi_r(q-1) = \sum_{i=1}^{q-1} \psi_r(i)
\end{equation}
After obtaining the optimal quantization level \(q_r^\star\), the optimal LT degree distribution is obtained by solving the inner problem defining \(\Phi_r(q_r^\star)\).

\begin{Remark}
\label{remark:p4}
Since all higher bitplanes than the current active one require error-free transmission, each additional quantization level incurs a strictly positive coding overhead, we select the first feasible level together with the inner LP solutions of its transmitted bitplanes, i.e.
\begin{equation}
	q_r^\star = q_{\min}
	= \min \left\{ q \in \{1,\dots,Q\} : P_{\max,q} > 0 \right\}. \notag
\end{equation}
\end{Remark}

\subsection{Client-Side Decoding with Conditional Priors}
An important component of the proposed system is client-side estimation of the distribution of the aggregated model weights $\tilde{\mathbf{w}}_r$ based on their local updated weights ${\mathbf{w}}^k_{r, E, l}$. 

Define the true model difference as 
\begin{equation}
	{\mathbf{z}}^k_{r} \triangleq \tilde{\mathbf{w}}_{r+1} - {\mathbf{w}}^k_{r, E},
\end{equation}
which is assumed to be another i.i.d. random variable with density $p({{z}}^k_{r})$.
The client predicts the distribution of aggregated weights via its local model and ${{z}}^k_{r}$, i.e.
\begin{align}
	p(\tilde{{w}}_{r+1, l} \mid {{w}}^k_{r, E, l}) &\triangleq p({w}^k_{r, E, l} + z), \quad z \sim p({{z}}^k_{r}) \notag\\
	&= p({{z}}^k_{r}) * \text{Dirac}({w}^k_{r, E, l}),
\end{align}
where `$*$' denotes the convolution operation.
It can be shown that the predicted difference between aggregated weights and local weights is always a noisy observation of the true difference, i.e.
\begin{equation} \label{eq66.1}
	\hat{\mathbf{z}}^k_{r, E} = {\mathbf{z}}^k_{r, E} + \delta^k_{r-1}.
\end{equation}
Despite the discrepancy, however, the clients can still approximate bitplane probabilities of $\tilde{\mathbf{w}}_{r+1}$. We introduce the following assumptions:
\begin{enumerate}
    \item $\delta^k_{r-1}$ and ${\mathbf{z}}^k_{r, E}$ are independent.
    \item $P_{r-1} \approx 0$ and $\Delta_{r} \approx \Delta_{r-1}$
\end{enumerate}

Define the client's prediction based on the noisy difference model as
\begin{align}
	\hat p(\tilde{{w}}_{r+1, l} \mid {{w}}^k_{r, E, l}) &\triangleq p({w}^k_{r, E, l} + z), \quad z \sim p(\hat{{z}}^k_{r}) \notag\\
	&= p(\hat{{z}}^k_{r}) * \text{Dirac}({w}^k_{r, E, l}),
\end{align}
With Eq. (\ref{eq66.1}) and assumption 1), we immediately have
\begin{equation}
	p(\hat{{z}}^k_{r}) = p({{z}}^k_{r}) * p(\delta^k_{r-1}),
\end{equation}
which leads to
\begin{equation}
	\hat p(\tilde{{w}}_{r+1, l} \mid {{w}}^k_{r, E, l}) = p(\tilde{{w}}_{r+1, l} \mid {{w}}^k_{r, E, l}) * p(\delta^k_{r-1}),
\end{equation}
consequently, we have
\begin{align}
\hat p\!\left(\tilde{w}_{r+1, l} \mid {{w}}^k_{r, E, l}\right) = \sum_{m=1}^{M} \pi^k_m\,
\mathcal{N}\!\left(\tilde{w}_{r+1, l} \mid \mu^k_m + w^k_{r,E,l},\;
(\sigma^k_m)^2\right).
\label{eq:gmm_shifted}
\end{align}
Recall the expression of $\delta_r$ in Lemma \ref{distortion_distribution}, under assumption 2), the convolution becomes
\begin{align}
	\hat p(\tilde{{w}}_{r+1, l} \mid {{w}}^k_{r, E, l}) \approx \frac{1}{\Delta_{r}}\int_{\tilde{{w}}_{r+1, l}-\frac{\Delta_{r-1}}{2}}^{\tilde{{w}}_{r+1, l}+\frac{\Delta_{r-1}}{2}}p(\tilde{{w}}_{r+1, l} \mid {{w}}^k_{r, E, l}) \,d{\tilde{{w}}_{r+1, l}}. \notag
\end{align}
Hence, we are able to estimate the per-bitplane conditional probabilities $\hat p(\{u_{r+1, l}\}_1^{q-1})$ and conditional entropies $H^k({u}^q_{r+1, l} \mid \{u_{r+1, l}\}_1^{q-1})$ at each client. 
The probability that the coefficient at position $j$ falls into the $n$-th bin is obtained by integrating the predicted density $\hat p(\tilde{{w}}_{r+1, l} \mid {{w}}^k_{r, E, l})$ over the cell, given by
\begin{align}
p(n \mid {{w}}^k_{r, E, l})&= \int_{T_{n,r}-(\frac{1}{2} + \xi_{r,l})\Delta_{r}}^{T_{n,r}+(\frac{1}{2} - \xi_{r,l})\Delta_{r}} p(\tilde{{w}}_{r+1, l} \mid {{w}}^k_{r, E, l}) d\tilde w_{r+1, l}, \notag\\
&\approx \Delta_{r} \hat p\!\left(T_{n,r} - \xi_{r,l}\Delta_{r} \mid {{w}}^k_{r, E, l}\right)
\label{eq:gmm_cell}
\end{align}
A quantization bin index $n \in \{0,\dots,2^{q_r}-1\}$ carries the $q-1$ most significant bits if and only if 
\[
\nu_{r+1, l}^{q-1}2^{q_r-q+1} \le n \le (\nu_{r+1, l}^{q-1}+1)2^{q_r-q+1} - 1.
\]
The per-bitplane prefix probability is then given by
\begin{align}
&p(\{u_{r+1, l}\}_1^{q-1}\mid {{w}}^k_{r, E, l}) \notag\\
&= \sum_{n = \nu_{r+1, l}^{q-1}2^{q_r-q+1}}^{(\nu_{r+1, l}^{q-1}+1)2^{q_r-q+1} - 1} p(n \mid {{w}}^k_{r, E, l}),
\qquad q = 1,\dots,Q
\label{eq:gmm_prefix}
\end{align}
Adopting the convention $p(u^{0}) = 1$, the probability of the $q$-th bit given the $q-1$ more significant bits is calculated with
\begin{align}
&p\!\left(u_{r+1, l}^q = b \mid \{u_{r+1, l}\}_1^{q-1}, {{w}}^k_{r, E, l}\right) \notag\\
&= \frac{p\!\left(\{u_{r+1, l}\}_1^{q-1},\, u_{r+1, l}^q = b \mid {{w}}^k_{r, E, l}\right)}
        {p\!\left(\{u_{r+1, l}\}_1^{q-1} \mid {{w}}^k_{r, E, l}\right)},
\quad b \in \{0,1\},
\label{eq:gmm_cond_bit}
\end{align}

For EXIT analysis, we empirically utilize the marginal bitplane distributions of $\tilde{\mathbf{w}}_{r+1}$ to calculate the prior probability for all systematic bits, which is directly counted with the quantized bits, i.e.
\begin{align} \label{counted_marginal}
&p\!\left(u_{r+1, l}^q = b \mid \{u_{r+1, l}\}_1^{q-1}, \mathbf{\tilde{w}}^k_{r+1}\right) \notag\\
&= \frac{\!\left|\{\{u_{r+1, l}\}_1^{q-1},\, u_{r+1, l}^q = b \}\right|}
        {\left|\{\{u_{r+1, l}\}_1^{q-1}\}\right|},
\quad b \in \{0,1\},
\end{align}

Since the conditional priors calculated from Eq. \eqref{eq:gmm_cond_bit} have lower entropy than the full marginal used for code design, we empirically employ an early-stopping mechanism where the client stops receiving more rateless symbols once the reliability of message symbols is greater than a certain threshold. 

The overall rateless downlink model broadcast procedure is summarized in Algorithm \ref{alg2}.
\RestyleAlgo{ruled}
\begin{algorithm}[!t] 
	\caption{Rateless Downlink Model Broadcast} \label{alg2}	
	\For{$r = 0,\ldots,R-1$}
	{
		\textbf{Server side:}\;
		Aggregate the client updates to obtain $\tilde{\mathbf{g}}_r$ and $\tilde{\mathbf{w}}_{r+1}$.\;
		Compute $\{\mathbf{u}_{r+1, q}\}_{q=1}^{q_r}$ with Eqs. (\ref{eq:quantization}) and (\ref{eq_binarize}).\;
		Compute $\bar D_r$ with Eq. (\ref{distortion_budget}).\;
		Compute marginal bitplane probabilities.
		\For{$k = 1,\ldots,K$}	
		{
			Fit the GMM parameters $\{\pi^k_m\}_{m=1}^{M}$, $\{\mu^k_m\}_{m=1}^{M}$, and $\{\sigma^k_m\}_{m=1}^{M}$ in Eq. (\ref{eq67}).\;
		}
		\For{$q = 1,\ldots,Q$}	
		{
			Calculate $P_r$ with Eq. (\ref{ber_target}).\;
			Calculate $H\!\left(u_{r+1, l}^q \mid \{u_{r+1, l}\}_1^{q-1}\right)$ with Eq. (\ref{counted_marginal}).\;
			\If{feasible $P_r > 0$}
				{
					$q_r \gets q$.\;
					Break.\;
				}
		}
		\For{$i = 1,\ldots,q_r-1$}
			{
				Solve \textbf{P2'} over the grid of $(\alpha_r, \bar y_r)$ for $\omega^i_r$.\;
			}
			Solve \textbf{P2} over the grid of $(\alpha_r, \bar y_r)$ for $\omega^{q_r}_r$.\;
		Transmit $\{\pi^k_m\}_{m=1}^{M}$, $\{\mu^k_m\}_{m=1}^{M}$, and $\{\sigma^k_m\}_{m=1}^{M}$ to the respective clients.\;
		Broadcast $\{\omega^i_r\}_{i=1}^{q_r}$ and $q_r$ to all clients.\;
		Perform rateless JSCC on $\tilde{\mathbf{w}}_{r+1}$ using $\{\omega^i_r\}_{i=1}^{q_r}$ and broadcast to all clients.\;

		\textbf{Client side:}\;
		\For{$k = 1,\ldots,K$}
		{
			Calculate bitplane conditional probabilities with Eq. (\ref{eq:gmm_cond_bit}).\;
			Reconstruct aggregated model ${\mathbf{w}}^k_{r+1, 0}$ with prior probability and BP decoding, subtracting the dither as in Eq. (\ref{eq:dedither}).\;
			Perform $E$ local training steps, upload the local update $\mathbf{g}^k_{r+1}$ to the server.\;
		}
	}
	\textbf{return} $\tilde{\mathbf{w}}_R$\;
\end{algorithm}

\section{Performance Evaluation}

In this section, we present numerical results of our proposed rateless downlink broadcast scheme to validate its robustness and bandwidth efficiency.

\subsection{Experimental Setup}

In simulations, we use the ResNet-18 model to evaluate the performance of the proposed rateless downlink broadcast scheme. We consider the CIFAR-10 dataset for use in both IID and non-IID data distributions. Specifically, for the non-IID setting, each client is assigned a primary class from which approximately 70\% of its samples are drawn, while the remaining samples are sourced from other classes. 
The simulation setup is summarized in Table \ref{table1}.

\begin{table}[!t]
\centering
\caption{{Experiment settings}}
\label{table1}
\begin{small}
	\begin{tabular}{|c|c|c|c|}
	\hline
	Parameter       & Value  & Parameter           & Value          \\ \hline
	$K$               & 3      & $d_{max}$           & 32             \\ \hline
	$E$               & 1      & $Q$                   & 14             \\ \hline
	$R$               & 10     & Max retransmissions & 2              \\ \hline
	$\eta$          & 0.0316 & $\bar y$            & $[0.3, 0.8]$ \\ \hline
	$p^k$           & 0.1    & $\alpha$            & $[4, 21]$    \\ \hline
	$q^k$           & 0.9    & $\epsilon_2$        & $1e-4$         \\ \hline
	$\tau$ & 200    & $D_0$               & $(5e-5)\mathcal{D}$         \\ \hline
	$d$               & 2000   & $\epsilon_1$        & $1e-5$         \\ \hline
	\end{tabular}
\end{small}
\end{table}

\subsubsection{Comparison Schemes}
We employ several benchmarks defined as follows.
\begin{itemize}
	\item LDPC + retransmission: a systematic $(3, 11)$ LDPC coding scheme where the entire codeword is transmitted. The erased blocks of each decoder are retransmitted to all clients.
	\item Mixed-timescale differential coding (MTDC) + LDPC + retransmission: we employ MTDC scheme \cite{hu2026mixed} as a compression method, where we set $\nu_0 = 16$, $\nu_1 = 14$, $\nu_2 = 12$, $\rho_1 = 10$, and $\rho_2 = 5$. The erased blocks of each decoder are retransmitted to all clients.
	\item Analog + retransmission: an uncoded analog transmission scheme where the aggregated weights are directly transmitted through the downlink channel \cite{9515709}. The erased blocks of each decoder are retransmitted to all clients. We set the analog power to $2$~dB.
	\item Proposed without conditional priors (CP): the proposed scheme with the conditional probability of model differences in Eq. (\ref{eq:gmm_cond_bit}) replaced by the marginal probability of aggregated weights in Eq. (\ref{counted_marginal}).
\end{itemize}

\subsection{Performance Comparison}
\begin{figure*}[t]
    \vspace{-29pt}
    \setlength{\belowcaptionskip}{-0.4cm}
	\centering
	\subfigure[]{\includegraphics[width=0.31\textwidth]{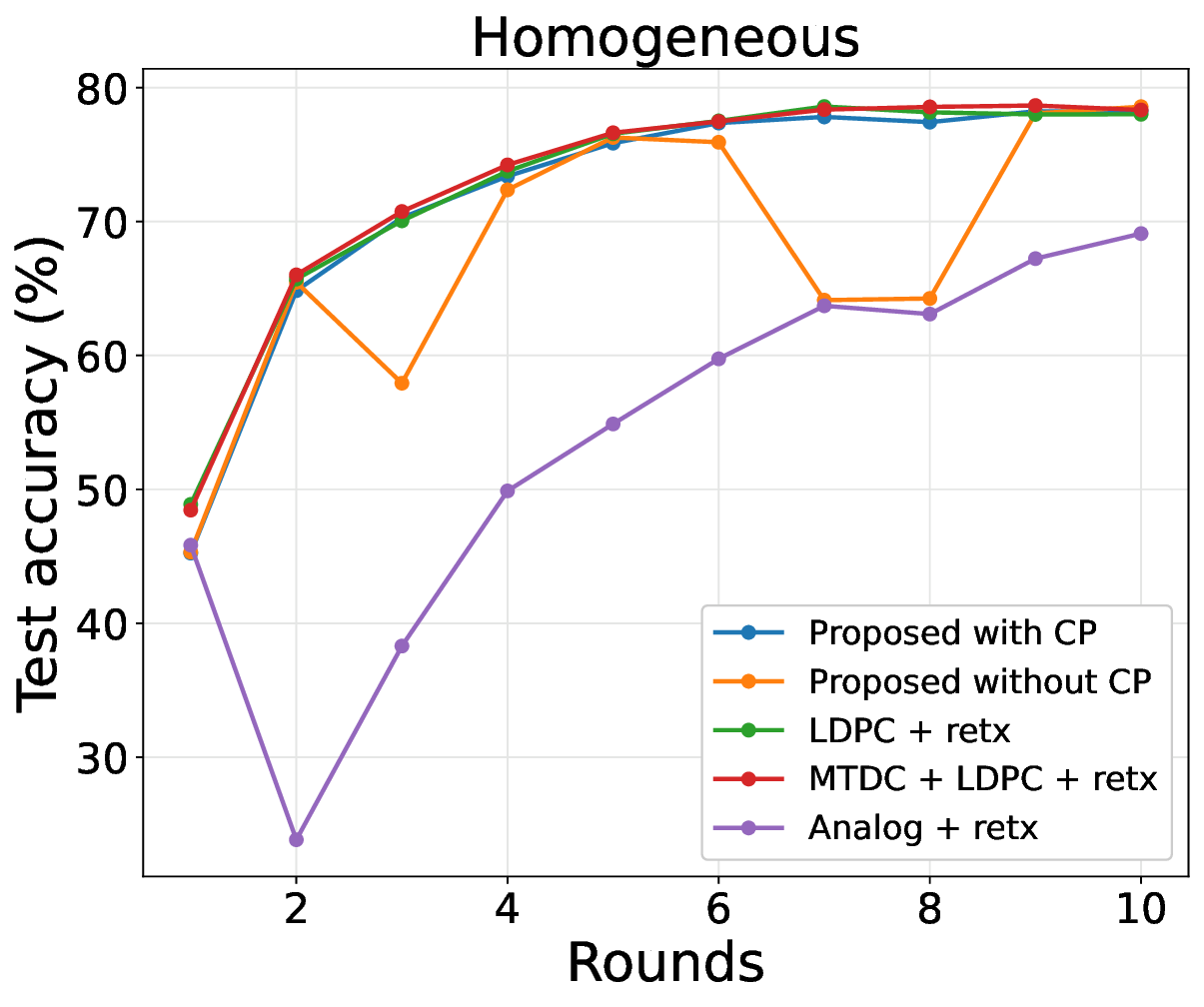}\label{fig2a}}
	\hfill
	\subfigure[]{\includegraphics[width=0.31\textwidth]{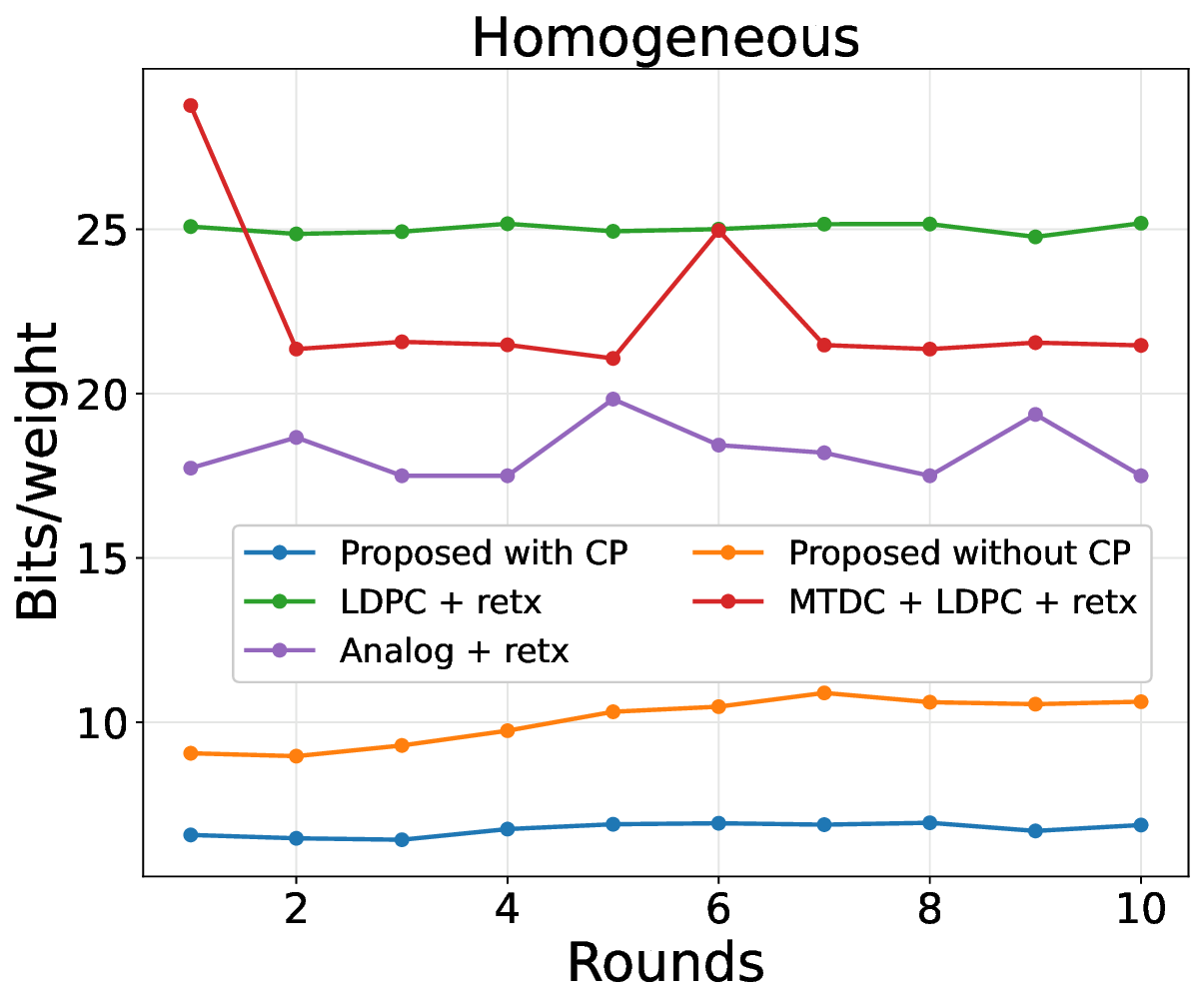}\label{fig3a}}
	\hfill
	\subfigure[]{\includegraphics[width=0.31\textwidth]{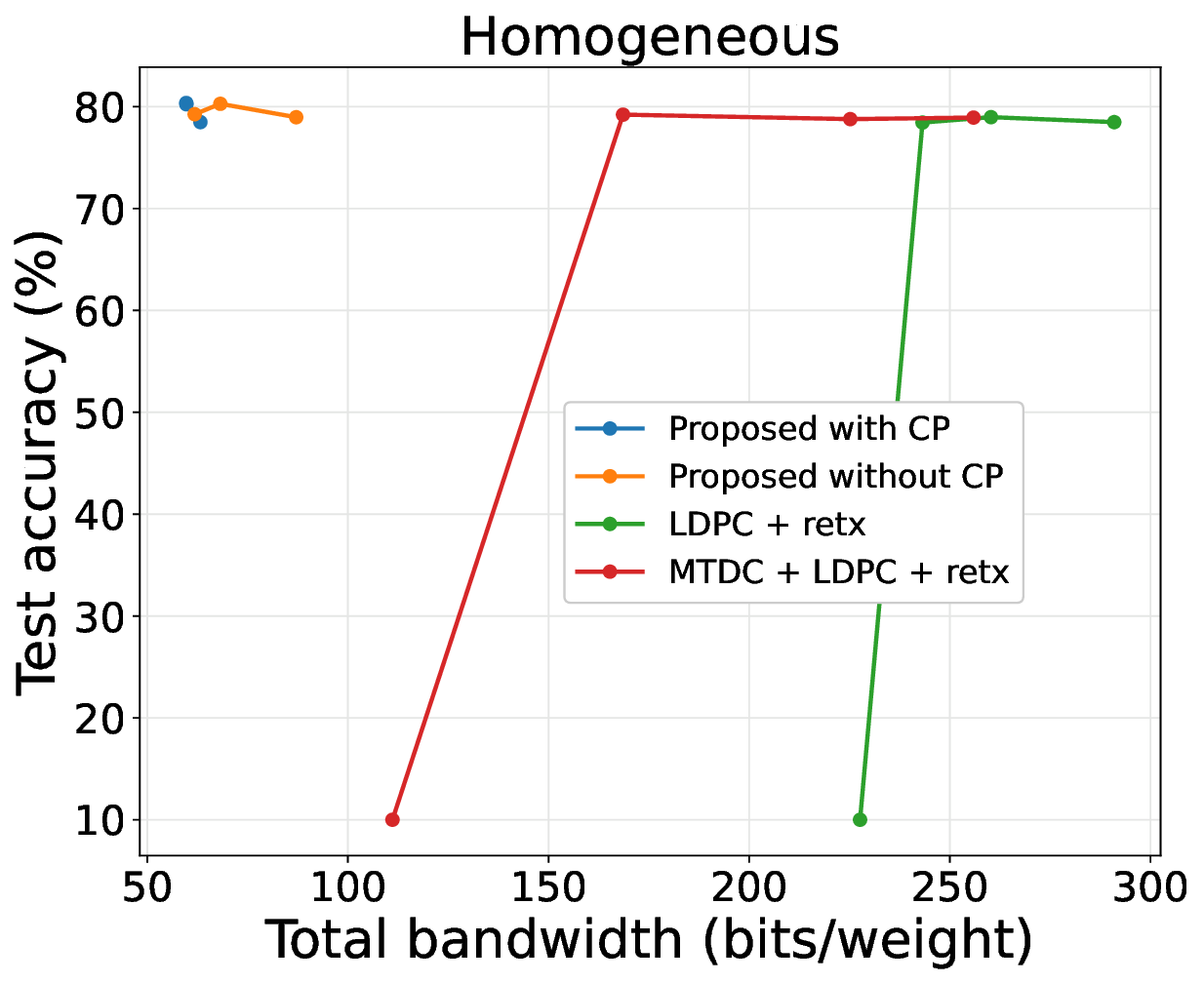}\label{fig5a}}
	\\[0pt]
	\subfigure[]{\includegraphics[width=0.31\textwidth]{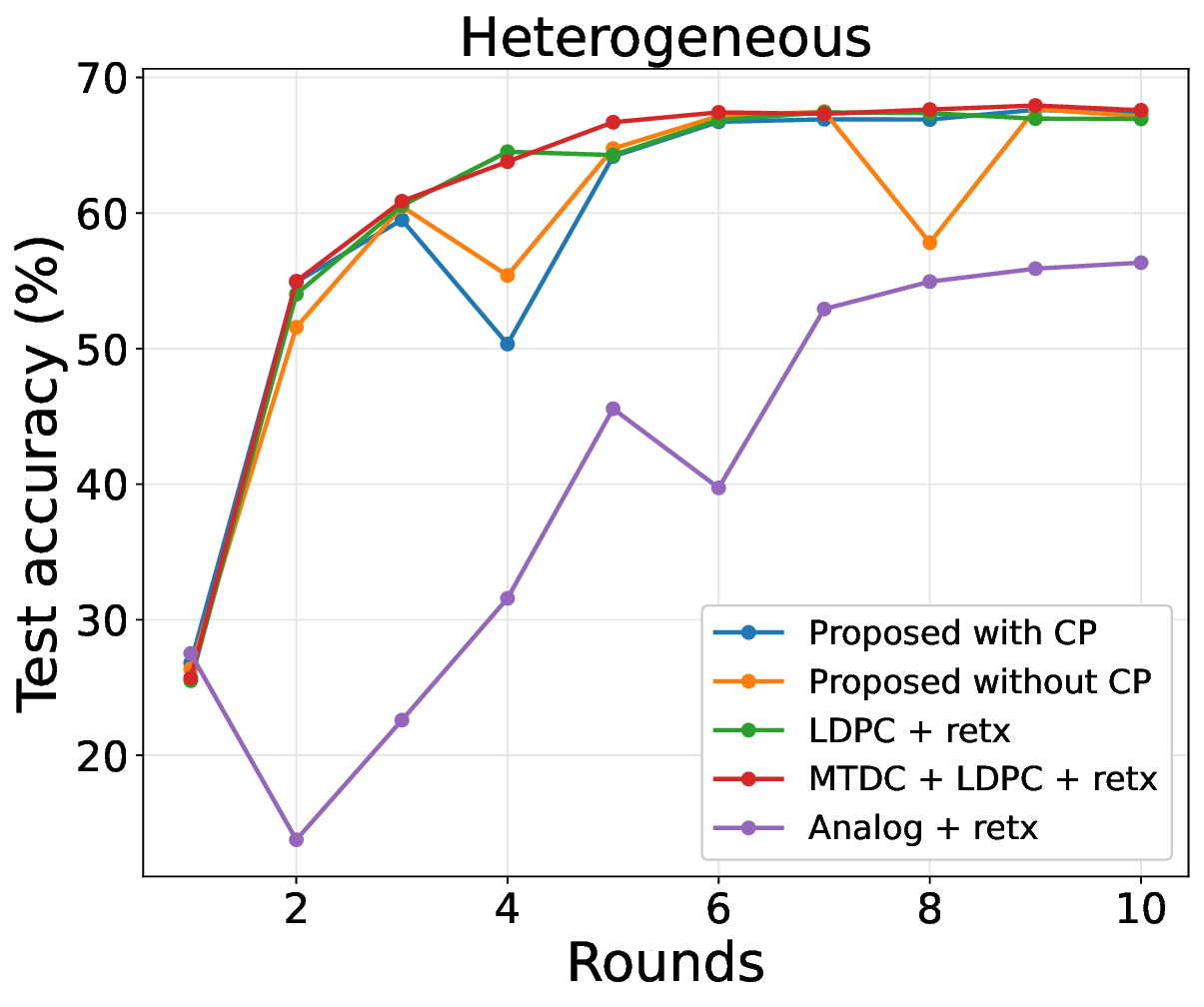}\label{fig2b}}
	\hfill
	\subfigure[]{\includegraphics[width=0.31\textwidth]{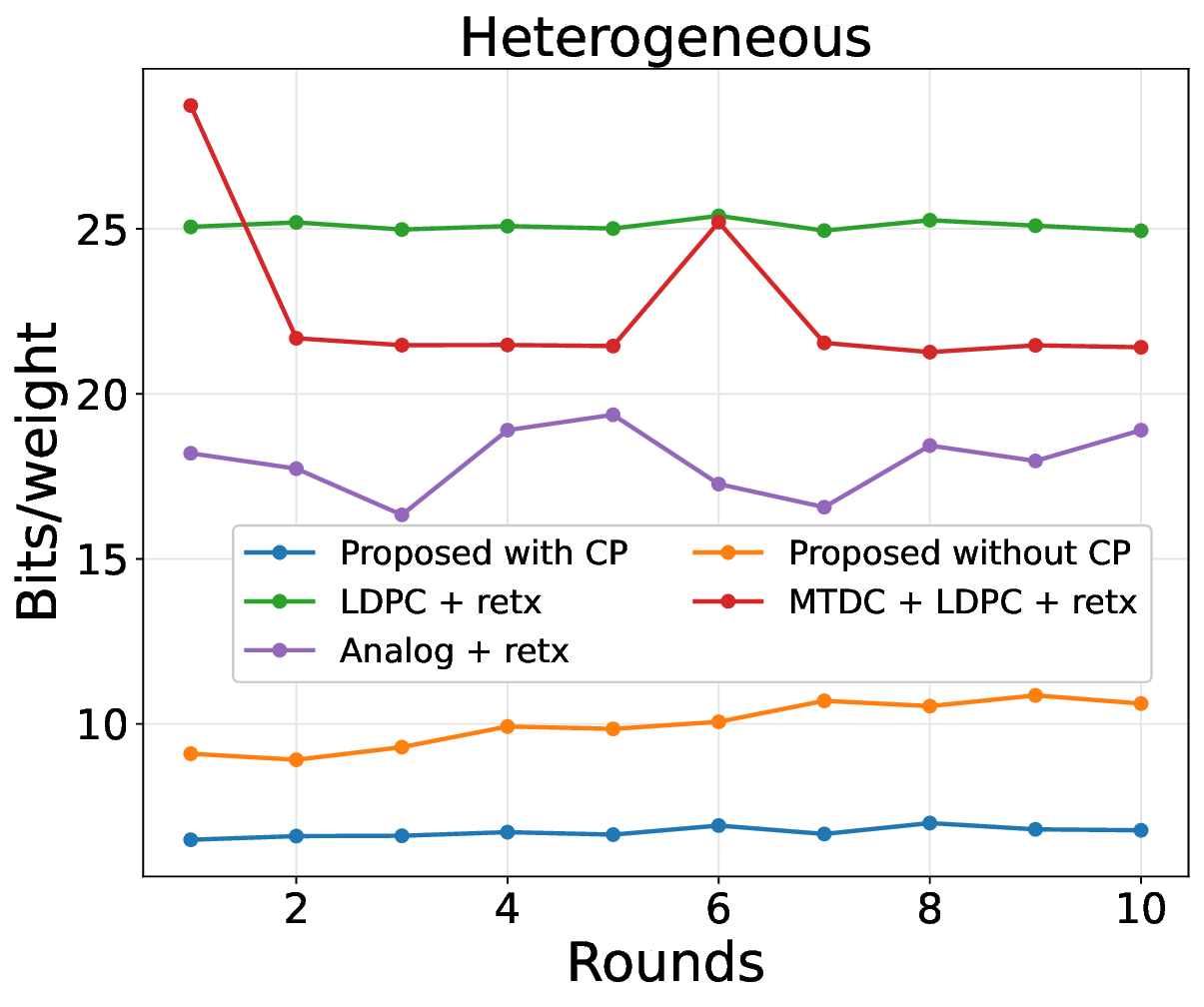}\label{fig3b}}
	\hfill
	\subfigure[]{\includegraphics[width=0.31\textwidth]{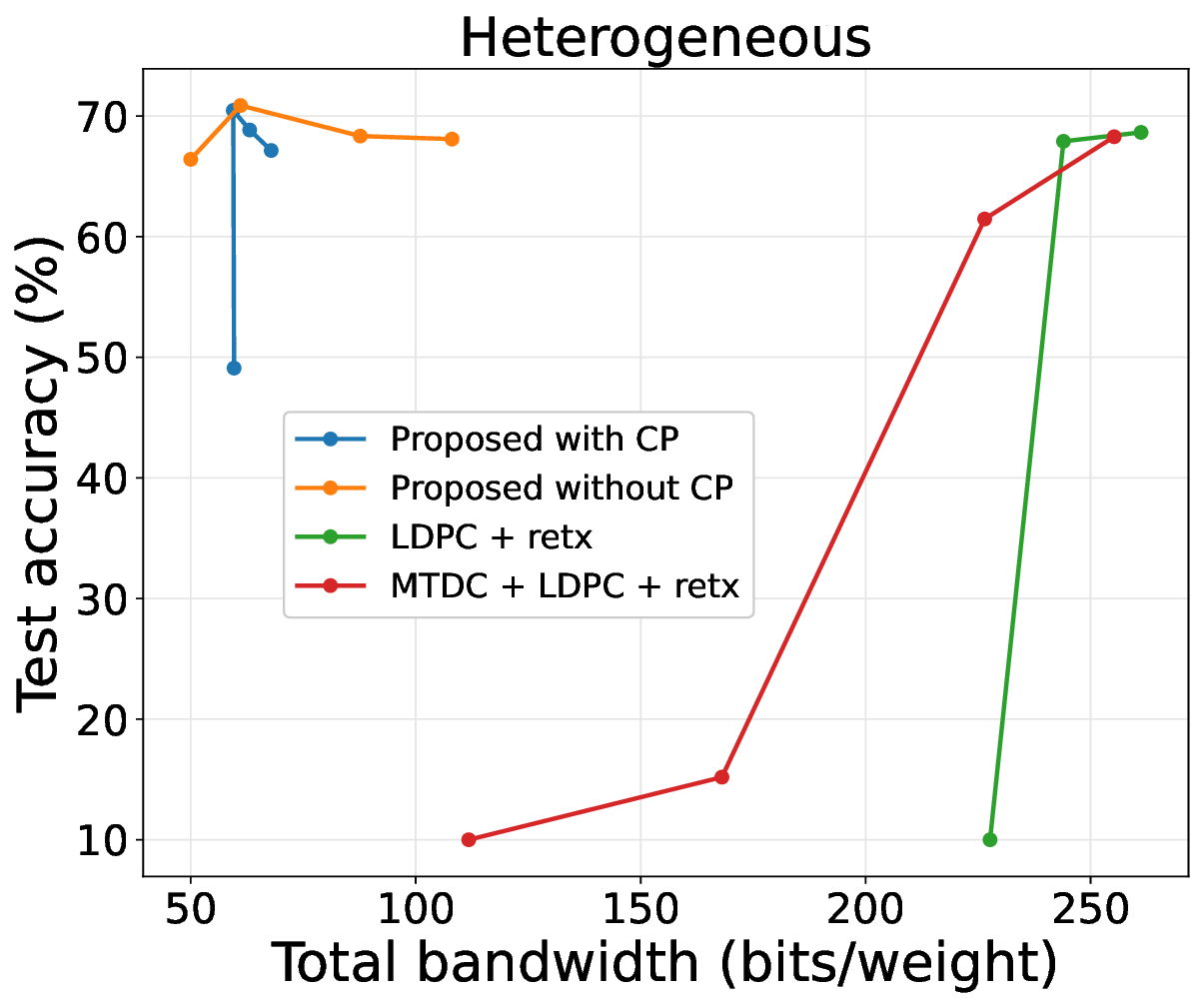}\label{fig5b}}
	\caption{Performance of the proposed rateless downlink broadcast scheme under homogeneous (top row) and heterogeneous (bottom row) client data: (a), (d) test accuracy versus communication rounds; (b), (e) per-round downlink bandwidth versus communication rounds; and (c), (f) test accuracy versus total downlink bandwidth.}
	\label{fig:performance}
\end{figure*}
Fig.~\ref{fig2a} and \ref{fig2b} compare the test accuracy versus communication rounds for different schemes under both homogeneous and heterogeneous data distributions across clients.
Fig.~\ref{fig2a} shows the results for homogeneous client data. The proposed Raptor$+$CP scheme, the Raptor$+$marginal baseline, the systematic LDPC benchmark and the MTDC$+$LDPC benchmark all achieve similar performance by the end of training: at the final round every method except the analog scheme attains between $78.0\%$ and $78.6\%$ accuracy. 
Fig.~\ref{fig2b} shows the results for heterogeneous client data. It can be seen that the test accuracy upon convergence becomes lower. Still, the proposed scheme reaches comparable accuracy with baselines, which is $67.3\%$ by the final round and the systematic LDPC and MTDC$+$LDPC benchmarks follow closely at $66.9\%$ and $67.6\%$.
Therefore, the results indicate that the proposed rateless design with conditional priors preserves the learning performance, and remains robust when the client data are non-IID.

Fig.~\ref{fig3a} and \ref{fig3b} compares the bandwidth consumption versus communication rounds for different schemes under both homogeneous and heterogeneous data distributions across clients.
Fig.~\ref{fig3a} shows the per-round downlink bandwidth in number of bits per model weight for the homogeneous case. 
All three non-raptor baseline methods suffer from high overhead due to inefficient retransmission of erased blocks to all clients.
On the other hand, the proposed scheme keeps the bandwidth nearly constant between $6.42$ and $6.94$ across all rounds, while the Raptor$+$marginal scheme grows from $9.06$ to $10.62$ as the quantization range widens with the diversification of the aggregated weight.
Fig.~\ref{fig3b} shows the same quantity under the heterogeneous partition, and the bandwidth profile of the schemes is essentially unchanged.

Fig.~\ref{fig5a} and \ref{fig5b} compares the final test accuracy against the total downlink bandwidth it consumes. 
Fig.~\ref{fig5a} shows the homogeneous case. The proposed scheme attains $78.5\%$--$80.4\%$ accuracy at total bandwidths of $59.7$--$63.3$ bits per weight, while the Raptor$+$marginal baseline spans $61.8$--$87.1$ bits per weight at $79.0\%$--$80.3\%$ accuracy; matching the best accuracy of the proposed scheme costs the latter about $68$ bits per weight. The MTDC$+$LDPC and systematic LDPC benchmarks only reach a comparable accuracy at $168.6$ and $243.2$ bits per weight, respectively, and their lower-bandwidth configurations fail to train, with the accuracy collapsing to $10\%$. 
Fig.~\ref{fig5b} shows the heterogeneous case, where the plateau drops to about $68\%$ for all schemes but the ordering is preserved: the proposed scheme attains $67.1\%$--$70.5\%$ at $59.5$--$67.9$ bits per weight and the Raptor$+$marginal baseline uses $61.1$--$108.1$ bits per weight for $68.1\%$--$70.9\%$, whereas the MTDC$+$LDPC and systematic LDPC benchmarks require at least $255.2$ and $244.0$ bits per weight and degrade to $61.5\%$ or fail entirely below that. 
We would like to note that for our proposed schemes both with and without conditional prior, greater bandwidth does not always correspond to better training performance, since it could also be the result of poor convergence of updates over rounds.

\begin{figure}[ht]
	\setlength{\belowcaptionskip}{-0.5cm}
	\centering
	\includegraphics[width=8.cm]{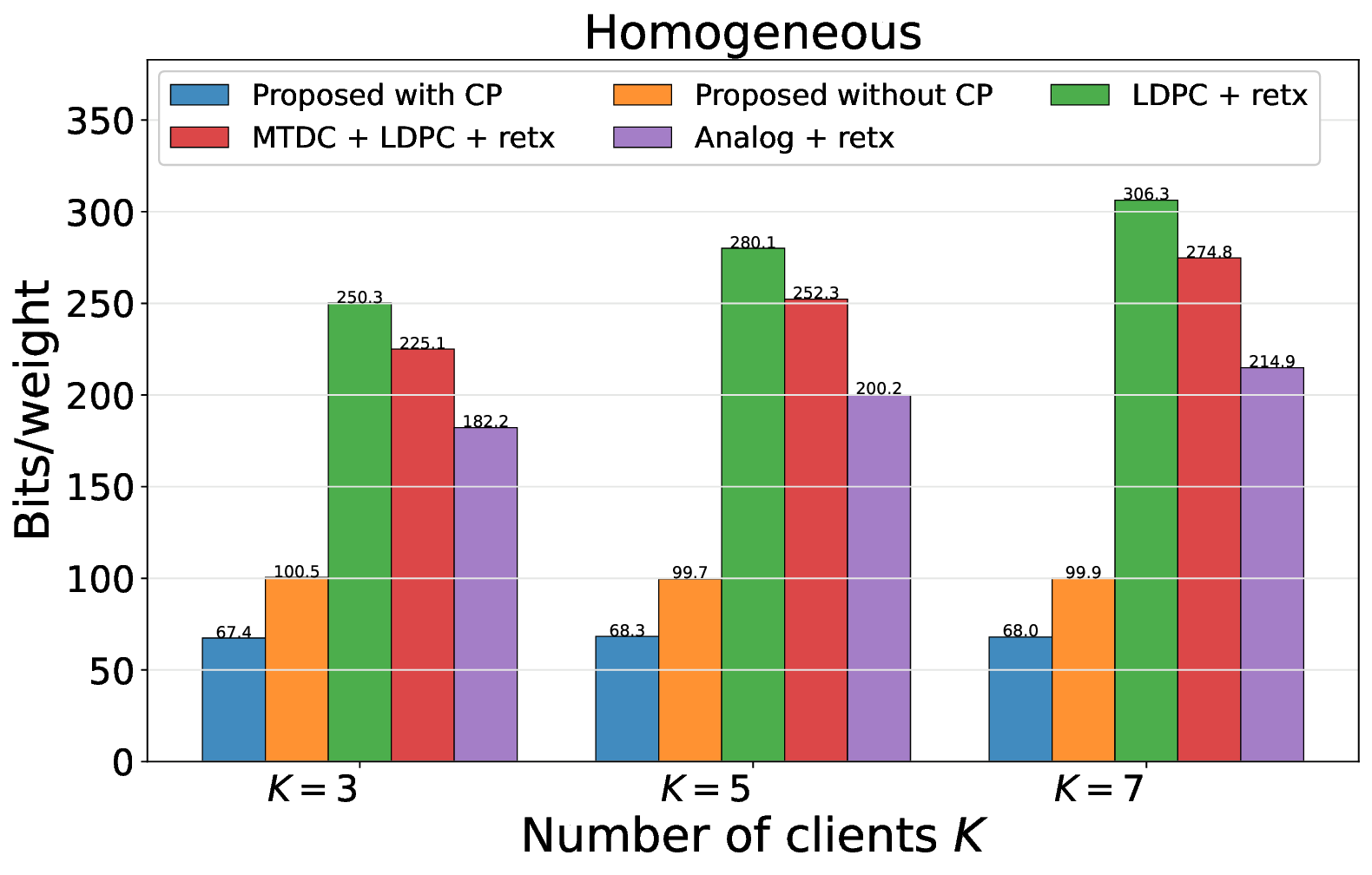}
	\caption{Total downlink bandwidth summed over the $R=10$ rounds for homogeneous client data.}
	\label{fig:total_bandwidth}
\end{figure}

Fig.~\ref{fig:total_bandwidth} compares the total bandwidth over the $10$ rounds for $K=3,5,7$. At $K=3$ the proposed scheme spends a total of $67.4$ channel uses per weight, against $100.5$ for Raptor$+$marginal, $182.2$ for the analog baseline, $225.1$ for MTDC$+$LDPC and $250.3$ for the systematic LDPC benchmark. 
Moreover, due to the rateless property of raptor codes, the total bandwidth of both our proposed scheme and the raptor$+$marginal scheme is affected very little by the number of clients, whereas the analog, MTDC$+$LDPC and systematic LDPC benchmarks all increase with the number of clients. Therefore, our proposed scheme is more advantageous when more clients are served. 



\section{Conclusion}
In this work, we proposed a rateless joint source-channel coding scheme for efficient downlink model broadcast in federated learning.
First, we investigated the convergence behavior of federated learning under noisy downlink broadcast, deriving an upper bound on the average squared gradient norm and a per-round distortion budget that links the transmission design to the learning performance. 
Then, we explored the design of adaptive quantization and systematic Raptor codes, where the multilayer EXIT analysis converts the distortion and decoding requirements into linear constraints. 
Finally, we exploited local weights from client-side models as prior information to aid decoding, and proposed an estimation method for bitplane probabilities based on noisy density information. 
The effectiveness of our method has been demonstrated through numerical experiments, allowing substantial reductions in downlink communication bandwidth while retaining the training accuracy under both homogeneous and heterogeneous data distributions. Future work may look into low-complexity estimation of conditional priors, as well as joint design of an uplink aggregation scheme.

\small
\bibliographystyle{IEEEtran}%
\bibliography{GradientBroadcast}

\clearpage
\normalsize
\appendix

\subsection{Proof of Lemma \ref{lemma1}}
\label{app:proof_lemma1}
\begin{proof}

Define
$\overline{\mathbf{h}}_{r,e}
    \triangleq
    \frac{1}{K}\sum_{k=1}^{K}
    \nabla f_k(\mathbf{w}_{r,e}^k).$
Assumptions~\ref{ass:unbiased_gradients} and
\ref{ass:bounded_local_variance} imply
\begin{equation}
    \mathbb{E}[\overline{\mathbf{g}}_{r,e}\mid\mathcal{F}_{r,e}]
    =\overline{\mathbf{h}}_{r,e}
\end{equation}
and
\begin{equation}
    \mathbb{E}\!\left[
    \|\overline{\mathbf{g}}_{r,e}-\overline{\mathbf{h}}_{r,e}\|^2
    \mid\mathcal{F}_{r,e}\right]
    \le\frac{\sigma_L^2}{K}.
    \label{eq:averaged_gradient_noise}
\end{equation}

By the smoothness of $f$ and \eqref{eq:virtual_server_update}, we have
\begin{align}
&\mathbb{E}\!\left[
 f(\widetilde{\mathbf{w}}_{r,e+1})\mid\mathcal{F}_{r,e}\right]
\notag\\
&\le f(\widetilde{\mathbf{w}}_{r,e})
-\eta\left\langle
\nabla f(\widetilde{\mathbf{w}}_{r,e}),
\overline{\mathbf{h}}_{r,e}\right\rangle
+\frac{L\eta^2}{2}
\mathbb{E}\!\left[
\|\overline{\mathbf{g}}_{r,e}\|^2
\mid\mathcal{F}_{r,e}\right].
\label{eq:conditional_smoothness_descent}
\end{align}
Using
\begin{equation}
    -\langle\mathbf{a},\mathbf{b}\rangle
    =-\frac{1}{2}\|\mathbf{a}\|^2
     -\frac{1}{2}\|\mathbf{b}\|^2
     +\frac{1}{2}\|\mathbf{a}-\mathbf{b}\|^2
\end{equation}
and
\begin{equation}
    \mathbb{E}[\|\overline{\mathbf{g}}_{r,e}\|^2
    \mid\mathcal{F}_{r,e}]
    \le\|\overline{\mathbf{h}}_{r,e}\|^2
    +\frac{\sigma_L^2}{K},
\end{equation}
we obtain, for $\eta L\le1$,
\begin{align}
\mathbb{E}f(\widetilde{\mathbf{w}}_{r,e+1})
&\le \mathbb{E}f(\widetilde{\mathbf{w}}_{r,e})
-\frac{\eta}{2}
 \mathbb{E}\|\nabla f(\widetilde{\mathbf{w}}_{r,e})\|^2
\notag\\
&\quad
+\frac{\eta}{2}
 \mathbb{E}\|\nabla f(\widetilde{\mathbf{w}}_{r,e})
 -\overline{\mathbf{h}}_{r,e}\|^2
+\frac{L\eta^2\sigma_L^2}{2K}.
\label{eq:descent_before_drift}
\end{align}
Furthermore, with Jensen's inequality, we have
\begin{align}\label{eq33.0}
\|\nabla f(\widetilde{\mathbf{w}}_{r,e})
-\overline{\mathbf{h}}_{r,e}\|^2 \le
\frac{1}{K}\sum_{k=1}^{K}\|\nabla f_k(\widetilde{\mathbf{w}}_{r,e})
-\nabla f_k(\mathbf{w}_{r,e}^k)\|^2.
\end{align}
Taking expectations over both sides and utilizing the smoothness of $f_k$ yields
\begin{align}\label{eq34.0}
\frac{1}{K}\sum_{k=1}^{K}\mathbb{E}\|\nabla f_k(\widetilde{\mathbf{w}}_{r,e})
-\nabla f_k(\mathbf{w}_{r,e}^k)\|^2
&\le \frac{L^2}{K}\sum_{k=1}^{K}
\mathbb{E}\|\mathbf{w}_{r,e}^k-
\widetilde{\mathbf{w}}_{r,e}\|^2.
\end{align}
Combining Eq. \eqref{eq33.0}, and \eqref{eq34.0}, we have
\begin{align}
\mathbb{E}\|\nabla f(\widetilde{\mathbf{w}}_{r,e})
    -\overline{\mathbf{h}}_{r,e}\|^2
    &\le \frac{L^2}{K}\sum_{k=1}^{K}
    \mathbb{E}\|\mathbf{w}_{r,e}^k-
    \widetilde{\mathbf{w}}_{r,e}\|^2.
\end{align}
Finally, substitution into \eqref{eq:descent_before_drift} yields
\begin{align}
\mathbb{E}f(\widetilde{\mathbf{w}}_{r,e+1})
&\le \mathbb{E}f(\widetilde{\mathbf{w}}_{r,e})
-\frac{\eta}{2}
 \mathbb{E}\|\nabla f(\widetilde{\mathbf{w}}_{r,e})\|^2
\notag\\
&\quad+\frac{\eta L^2}{2K}\sum_{k=1}^{K}
    \mathbb{E}\|\mathbf{w}_{r,e}^k-
    \widetilde{\mathbf{w}}_{r,e}\|^2
+\frac{L\eta^2\sigma_L^2}{2K}.
\end{align}
which completes the proof.
\end{proof}

\subsection{Proof of Lemma \ref{lemma2}}
\label{app:proof_lemma2}
\begin{proof}
An extension to Jensen's inequality states that for any vectors $\mathbf{a}_1,\dots,\mathbf{a}_n$,
\begin{equation} \label{ext_jensen}
	\Bigg|\Bigg|\sum_{i=1}^{n}\mathbf{a}_i\Bigg|\Bigg|^2
	\le n\sum_{i=1}^{n}\|\mathbf{a}_i\|^2.
\end{equation}
This inequality will be used repeatedly in the following derivation.

By definition, we have
\begin{equation}
    \mathbf{w}_{r,e}^k-
    \widetilde{\mathbf{w}}_{r,e}
    =\bm{\delta}_{r-1}^k
    -\eta\sum_{s=0}^{e-1}
    (\mathbf{g}_{r,s}^k-\overline{\mathbf{g}}_{r,s}).
\end{equation}
Taking expectations on both sides and applying Eq. \eqref{ext_jensen}, we have 
\begin{align}
&\frac{1}{K}\sum_{k=1}^{K}\mathbb{E}\|\mathbf{w}_{r,e}^k-
    \widetilde{\mathbf{w}}_{r,e}\|^2 \notag\\
&\le\frac{2}{K}\sum_{k=1}^{K} \mathbb{E}\|\bm{\delta}_{r-1}^k\|^2
+\frac{2\eta^2}{K}\sum_{k=1}^{K}\mathbb{E}\Bigg|\Bigg|\sum_{s=0}^{e-1}
(\mathbf{g}_{r,s}^k-
\overline{\mathbf{g}}_{r,s})\Bigg|\Bigg|^2 \notag\\
&\le\frac{2}{K}\sum_{k=1}^{K} \mathbb{E}\|\bm{\delta}_{r-1}^k\|^2
+2\eta^2e\sum_{s=0}^{e-1}
\frac{1}{K}\sum_{k=1}^{K}
\mathbb{E}\|\mathbf{g}_{r,s}^k-
\overline{\mathbf{g}}_{r,s}\|^2.
\label{eq:drift_initial_bound}
\end{align}
Moreover, 
we have
\begin{align}
\frac{1}{K}\sum_{k=1}^{K}
\mathbb{E}\|\mathbf{g}_{r,s}^k-
\overline{\mathbf{g}}_{r,s}\|^2
&\le\frac{1}{K}\sum_{k=1}^{K}
\mathbb{E}\|\nabla f_k(\mathbf{w}_{r,s}^k)\|^2
+\sigma_L^2.
\label{eq:gradient_dispersion_bound}
\end{align}
For each client, we decompose its local gradient with
\begin{align}
	\nabla f_k(\mathbf{w}_{r,s}^k) &= \left[\nabla f_k(\mathbf{w}_{r,s}^k) - \nabla f_k(\widetilde{\mathbf{w}}_{r,s})\right] \notag\\
	&+ \left[\nabla f_k(\widetilde{\mathbf{w}}_{r,s}) - \nabla f(\widetilde{\mathbf{w}}_{r,s})\right]
	+ \nabla f(\widetilde{\mathbf{w}}_{r,s}).
\end{align}
Utilizing Eq. \eqref{ext_jensen}, 
Assumptions~\ref{ass:smoothness} and
\ref{ass:gradient_dissimilarity}, we have
\begin{align}
&\frac{1}{K}\sum_{k=1}^{K}
\mathbb{E}\|\nabla f_k(\mathbf{w}_{r,s}^k)\|^2
\notag\\
& \le \frac{3}{K}\sum_{k=1}^{K}
    \mathbb{E}\|\nabla f_k(\mathbf{w}_{r,s}^k) - \nabla f_k(\widetilde{\mathbf{w}}_{r,s})\|^2 \notag\\
&+\frac{3}{K}\sum_{k=1}^{K}\|\nabla f_k(\widetilde{\mathbf{w}}_{r,s}) - \nabla f(\widetilde{\mathbf{w}}_{r,s})\|^2
+3\mathbb{E}\|\nabla f(\widetilde{\mathbf{w}}_{r,s})\|^2 \notag\\
&\le\frac{3L^2}{K}\sum_{k=1}^{K}
    \mathbb{E}\|\mathbf{w}_{r,s}^k-
    \widetilde{\mathbf{w}}_{r,s}\|^2+3\sigma_G^2
+3\mathbb{E}\|\nabla f(\widetilde{\mathbf{w}}_{r,s})\|^2.
\label{eq:local_gradient_moment_bound}
\end{align}
Combining \eqref{eq:drift_initial_bound}--
\eqref{eq:local_gradient_moment_bound}, we have
\begin{align}
&\frac{1}{K}\sum_{k=1}^{K}
    \mathbb{E}\|\mathbf{w}_{r,e}^k-
    \widetilde{\mathbf{w}}_{r,e}\|^2 \notag\\
&\le\frac{2}{K}\sum_{k=1}^{K} \mathbb{E}\|\bm{\delta}_{r-1}^k\|^2+\frac{6\eta^2eL^2}{K}\sum_{s=0}^{e-1}\sum_{k=1}^{K}
    \mathbb{E}\|\mathbf{w}_{r,s}^k-
    \widetilde{\mathbf{w}}_{r,s}\|^2,
\notag\\
&\quad+2\eta^2e\sum_{s=0}^{e-1}
\left[
3\mathbb{E}\|\nabla f(\widetilde{\mathbf{w}}_{r,s})\|^2
+3\sigma_G^2+\sigma_L^2
\right].
\label{eq:recursive_drift_bound}
\end{align}
Let
\[
A_r^{\max}\triangleq\max_{0\le e\le E}\frac{1}{K}\sum_{k=1}^{K}
    \mathbb{E}\|\mathbf{w}_{r,e}^k-
    \widetilde{\mathbf{w}}_{r,e}\|^2.
\]
Since $e\le E$, Eq. \eqref{eq:recursive_drift_bound} yields
\begin{align}
A_r^{\max}
&\le\frac{2}{K}\sum_{k=1}^{K} \mathbb{E}\|\bm{\delta}_{r-1}^k\|^2+6\eta^2L^2E^2A_r^{\max}
\notag\\
&\quad+6\eta^2E\sum_{s=0}^{E-1}
\mathbb{E}\|\nabla f(\widetilde{\mathbf{w}}_{r,s})\|^2
+2\eta^2E^2(3\sigma_G^2+\sigma_L^2).
\label{eq:max_drift_before_absorption}
\end{align}
If $6\eta^2L^2E^2\le1/2$, then
\begin{align}
A_r^{\max}
&\le\frac{4}{K}\sum_{k=1}^{K} \mathbb{E}\|\bm{\delta}_{r-1}^k\|^2
+12\eta^2E\sum_{s=0}^{E-1}
\mathbb{E}\|\nabla f(\widetilde{\mathbf{w}}_{r,s})\|^2
\notag\\
&\quad+4\eta^2E^2(3\sigma_G^2+\sigma_L^2).
\label{eq:max_drift_bound}
\end{align}
Therefore,
\begin{align}
&\sum_{e=0}^{E-1}\frac{1}{K}\sum_{k=1}^{K}
    \mathbb{E}\|\mathbf{w}_{r,e}^k-
    \widetilde{\mathbf{w}}_{r,e}\|^2 \notag\\
&\le\frac{4E}{K}\sum_{k=1}^{K} \mathbb{E}\|\bm{\delta}_{r-1}^k\|^2
+12\eta^2E^2\sum_{e=0}^{E-1}
\mathbb{E}\|\nabla f(\widetilde{\mathbf{w}}_{r,e})\|^2
\notag\\
&\quad+4\eta^2E^3(3\sigma_G^2+\sigma_L^2),
\end{align}
which completes the proof.
\end{proof}

\subsection{Proof of Theorem \ref{Theorem1}}
\label{app:proof_theorem1}
\begin{proof}
Summing \eqref{eq:one_step_descent_corrected} over
$e=0,\ldots,E-1$ and using
$\widetilde{\mathbf{w}}_{r,E}=\widetilde{\mathbf{w}}_{r+1}$ gives
\begin{align}
&\frac{\eta}{2}\sum_{e=0}^{E-1}
\mathbb{E}\|\nabla f(\widetilde{\mathbf{w}}_{r,e})\|^2
\le
\mathbb{E}f(\widetilde{\mathbf{w}}_r)
-\mathbb{E}f(\widetilde{\mathbf{w}}_{r+1}) \notag\\
&+\frac{\eta L^2}{2}\sum_{e=0}^{E-1}\frac{1}{K}\sum_{k=1}^{K}
    \mathbb{E}\|\mathbf{w}_{r,e}^k-
    \widetilde{\mathbf{w}}_{r,e}\|^2
+\frac{L\eta^2E\sigma_L^2}{2K}.
\label{eq:round_descent_before_substitution}
\end{align}
Substituting \eqref{eq:summed_drift_bound} yields
\begin{align}
&\left(\frac{\eta}{2}-6\eta^3L^2E^2\right)
\sum_{e=0}^{E-1}
\mathbb{E}\|\nabla f(\widetilde{\mathbf{w}}_{r,e})\|^2
\notag\\
&\le
\mathbb{E}f(\widetilde{\mathbf{w}}_r)
-\mathbb{E}f(\widetilde{\mathbf{w}}_{r+1})
+2\eta L^2E\frac{1}{K}\sum_{k=1}^{K} \mathbb{E}\|\bm{\delta}_{r-1}^k\|^2
\notag\\
&\quad+2\eta^3L^2E^3(3\sigma_G^2+\sigma_L^2)
+\frac{L\eta^2E\sigma_L^2}{2K}.
\label{eq:round_descent_after_substitution}
\end{align}
Condition \eqref{eq:theorem_stepsize_condition} implies
$6\eta^2L^2E^2\le1/4$, 
which leads to
\begin{align} \label{eq53}
&\frac{\eta}{4}\sum_{e=0}^{E-1}
\mathbb{E}\|\nabla f(\widetilde{\mathbf{w}}_{r,e})\|^2
\notag\\
&\le
\mathbb{E}f(\widetilde{\mathbf{w}}_r)
-\mathbb{E}f(\widetilde{\mathbf{w}}_{r+1})
+2\eta L^2E\frac{1}{K}\sum_{k=1}^{K} \mathbb{E}\|\bm{\delta}_{r-1}^k\|^2
\notag\\
&\quad+2\eta^3L^2E^3(3\sigma_G^2+\sigma_L^2)
+\frac{L\eta^2E\sigma_L^2}{2K}.
\end{align}
Summing Eq. \eqref{eq53} over $r=0,\ldots,R-1$, applying
$f(\widetilde{\mathbf{w}}_R)\ge f^\star$, and dividing by
$\eta RE/4$ proves \eqref{eq:theorem1_corrected_bound}.
\end{proof}

\end{document}